\documentclass{article}
\usepackage[utf8]{inputenc}
\usepackage{amsfonts}
\usepackage{amsmath}
\usepackage[percent]{overpic}
\usepackage{verbatim}
\usepackage{xcolor}
\usepackage{mdframed}
\usepackage{makecell}
\usepackage{xstring}
\usepackage{multirow}
\usepackage{anyfontsize}
\usepackage{tabularx}
\usepackage{wrapfig}
\usepackage{datatool}
\DTLsetseparator{:}
\usepackage{xifthen}
\usepackage{listofitems}
\usepackage{floatrow}
\usepackage[affil-it]{authblk}
\usepackage{ifthen}
\usepackage{comment}
\usepackage{bbm}
\usepackage[many]{tcolorbox}
\usepackage[euler]{textgreek}
\usepackage{algorithm}
\usepackage{textcomp}
\usepackage[noend]{algpseudocode}
\usepackage{url}
\usepackage{subcaption}
\usepackage{graphicx}
\usepackage{rotating}
\usepackage{standalone}

\makeatletter
\def\BState{\State\hskip-\ALG@thistlm}
\makeatother

\usepackage{transparent}
\usepackage[margin=1in]{geometry}

\usepackage{lineno}

\title{Deep Learning for Day Forecasts  from Sparse Observations}

\author[*1]{Marcin Andrychowicz}
\author[*1]{Lasse Espeholt}
\author[*1]{Di Li}
\author[2]{Samier Merchant}
\author[2]{Alexander Merose}
\author[2]{Fred Zyda}
\author[2]{Shreya Agrawal}
\author[*1]{Nal Kalchbrenner}

\affil[1]{Google DeepMind}
\affil[2]{Google Research}
\affil[*]{equal contribution}

\date{June 2023}

\usepackage[square,sort,comma,numbers]{natbib}
\usepackage{graphicx}
\usepackage{tikz}
 
\usepackage{todonotes}
\usepackage{subcaption}
\usepackage{caption}
\usepackage{booktabs}
\usepackage{float}
\usepackage{diagbox}

\usepackage{makecell}

\usepackage{xr-hyper}
\usepackage{hyperref}
\makeatletter

\newcommand*{\addFileDependency}[1]{%
\typeout{(#1)}%
\@addtofilelist{#1}
\IfFileExists{#1}{}{\typeout{No file #1.}}
}\makeatother

\newcommand{\myexternaldocument}[1]{%
\externaldocument{#1}%
\addFileDependency{#1.tex}%
\addFileDependency{#1.aux}%
}

\myexternaldocument{supplement}

\begin{document}
\maketitle

\begin{abstract}
Deep neural networks offer an alternative paradigm for modeling weather conditions. The ability of neural models to make a prediction in less than a second once the data is available and to do so with very high temporal and spatial resolution, and the ability to learn directly from atmospheric observations, are just some of these models' unique advantages. Neural models trained using atmospheric observations, the highest fidelity and lowest latency data, have to date achieved good performance only up to twelve hours of lead time when compared with state-of-the-art probabilistic Numerical Weather Prediction models and only for the sole variable of precipitation.
In this paper, we present MetNet-3 that extends significantly both the lead time range and the variables that an observation based neural model can predict well. MetNet-3 learns from both dense and sparse data sensors and makes predictions up to 24 hours ahead for precipitation, wind,
temperature and dew point. MetNet-3 introduces a key densification technique that implicitly captures data assimilation and produces spatially dense forecasts in spite of the network  training on extremely sparse targets.
MetNet-3 has a high temporal and spatial resolution of, respectively, up to 2 minutes and 1~km as well as a low operational latency.
We find that MetNet-3 is able to outperform the best single- and multi-member NWPs such as HRRR and ENS over the CONUS region for up to 24 hours ahead,
setting a new performance milestone for observation based neural models.
MetNet-3 is operational and its forecasts are served in Google Search in conjunction with other models.

\end{abstract}

\section{Introduction}
Physics based Numerical Weather Prediction (NWP) models currently drive the main forecasts that are available worldwide.
These systems collect and process a large number of sparse and dense sources of observations of the atmosphere into an initial dense atmospheric representation via a process called data assimilation, which they then roll out into the future using physical laws approximations. The  forward simulation is an expensive process that requires
thousands of CPU hours just to make a single forecast for hours or days ahead. The spatial and temporal resolutions of the forecasts must be kept relatively low as they dramatically affect the computational cost of the simulation. %

Weather models based on neural networks that use direct atmospheric observations for training offer an alternative modeling paradigm. Once the observations are available neural models have a prediction latency that is in the order of seconds. The forecast spatial resolution of the model has limited impact on computational cost that enable forecasts of one kilometer spatial resolution or higher and a very high temporal resolution in the order of minutes. Neural models can also learn atmospheric phenomena directly from the observations that capture them. This removes the need to explicitly describe a weather phenomenon using complex physics and makes it possible to model phenomena for which the physics is not well understood or that go beyond the usual domain of weather.

These advantageous properties make neural models a strong contender for an alternative paradigm for atmospheric modeling. However,  high-resolution neural weather models have only been shown to perform well up to twelve hours of lead time and on the sole domain of precipitation~\cite{Espeholt2022}. Identifying, processing and packaging for neural training the many sources of observational data that are needed to capture sufficient atmospheric information in the first place is an inordinate engineering challenge. Observational data sources come from a large number of providers with differing formats, have different spatial and temporal resolutions, and different degrees of sparsity ranging from individual points, like those from weather stations, to dense geospatial images like the observations that arise from ground-based radars and orbiting satellites. The widely different degrees of sparsity represent a novel machine learning challenge in and of themselves, as the model is expected to learn from sparse data, but produce a dense forecast.

\begin{figure}[t]
    \begin{center}
    \begin{minipage}[b][][b]{.2\textwidth}
    \centering
    \scriptsize
    \textbf{(a) Dropout training}\
    \includegraphics[width=\textwidth]{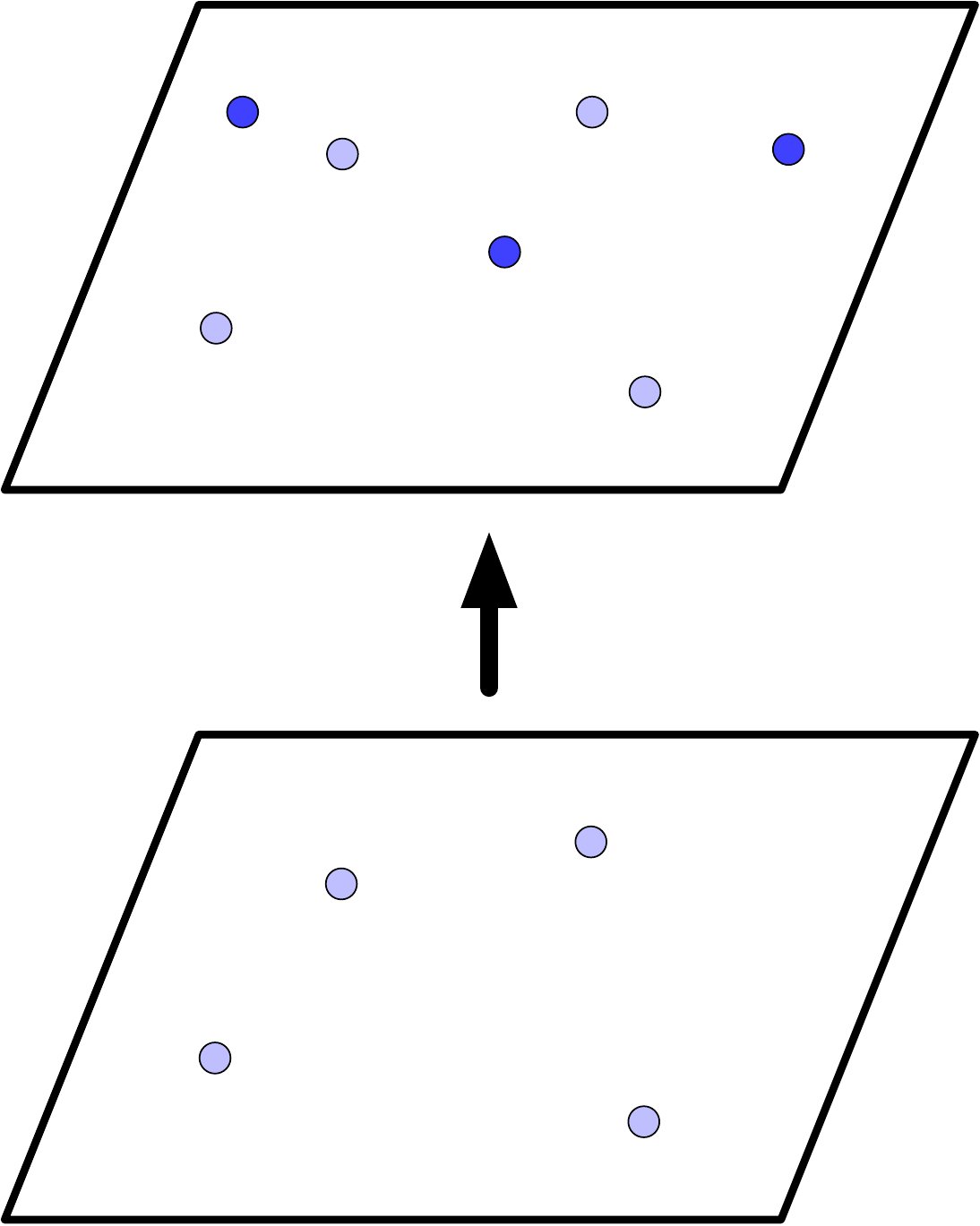}\
    \end{minipage}%
    \hspace{.5cm}
    \begin{minipage}[b][][b]{.2\textwidth}
    \centering
    \scriptsize
    \textbf{(b) Evaluation\\ of generalization}\
    \includegraphics[width=\textwidth]{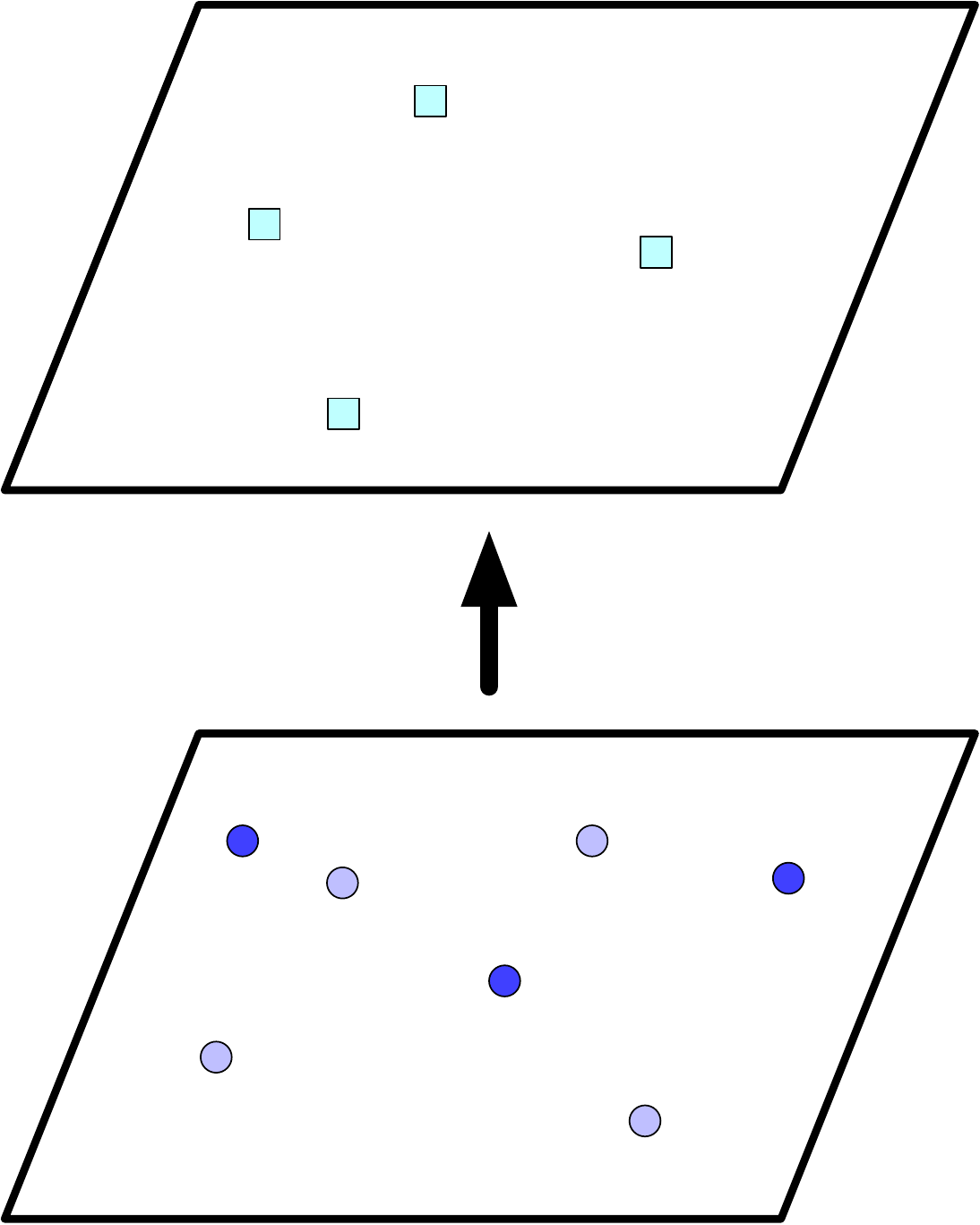}\
    \end{minipage}%
    \hspace{.5cm}
    \begin{minipage}[b][][b]{.2\textwidth}
    \centering
    \scriptsize
    \textbf{(c) Hyperlocal evaluation}\
    \includegraphics[width=\textwidth]{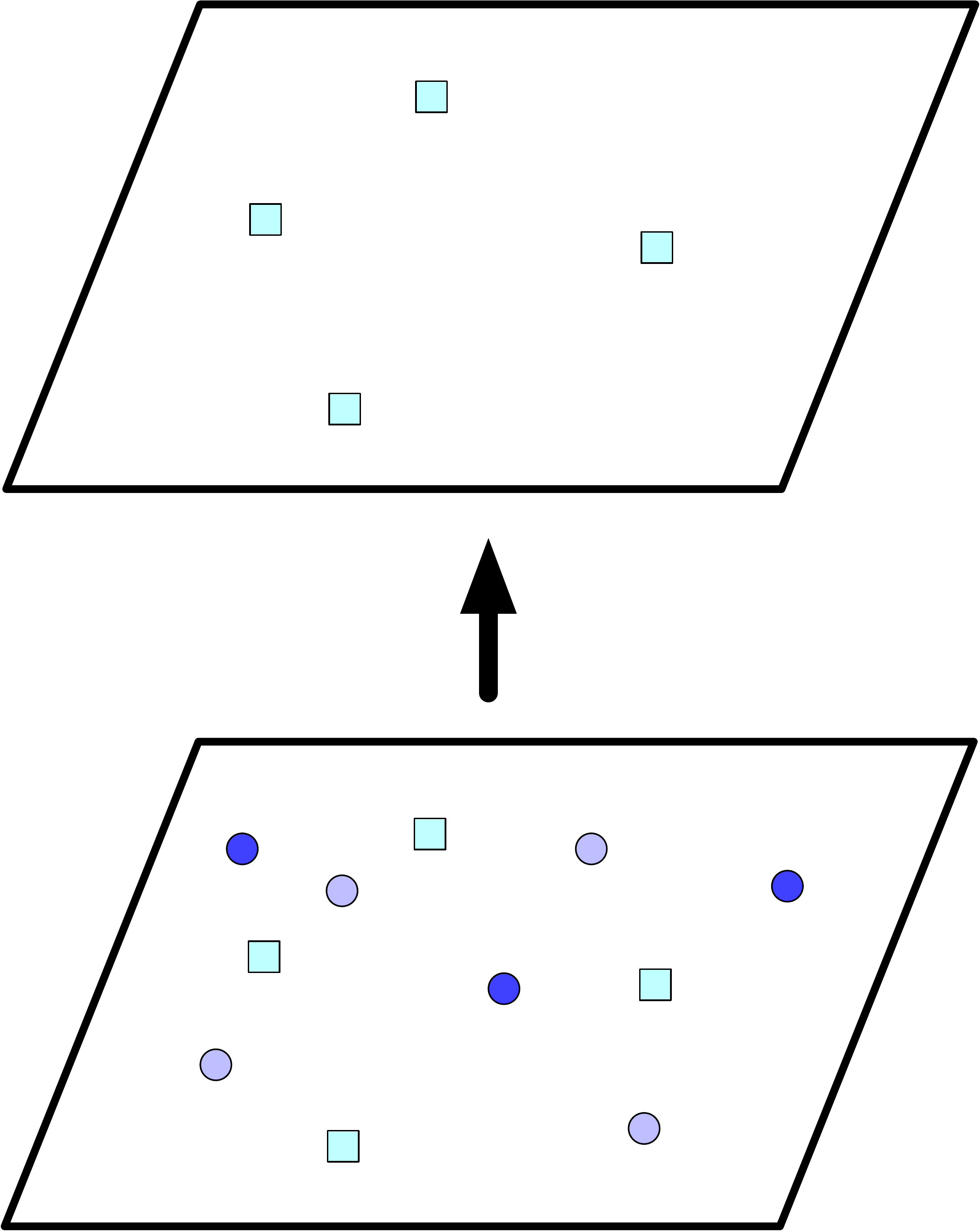}\
    \end{minipage}%
    \hspace{.5cm}
    \begin{minipage}[b][][b]{.2\textwidth}
    \centering
    \scriptsize
    \textbf{(d) Densified forecast}\
    \includegraphics[width=\textwidth]{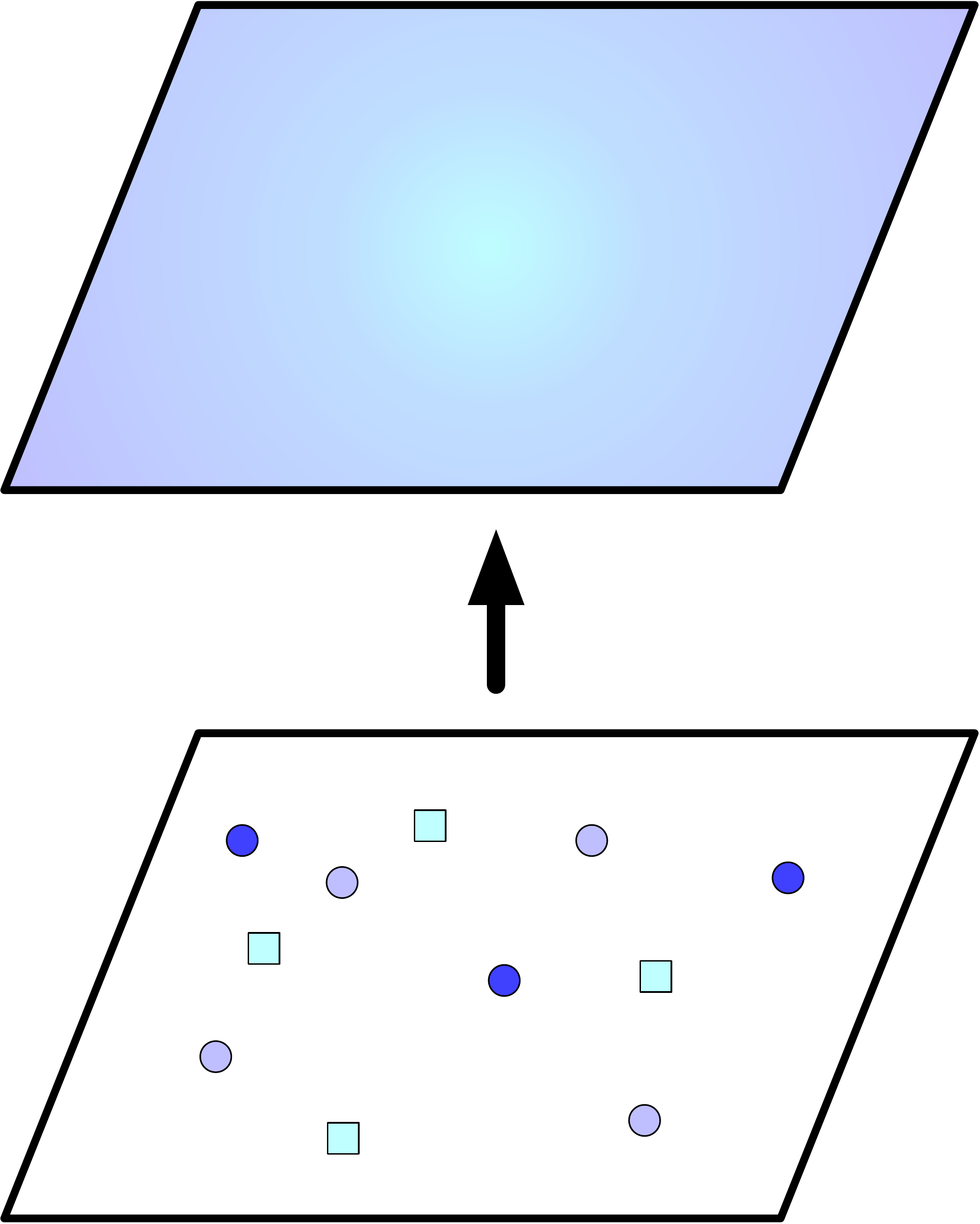}\
    \end{minipage}
\caption{Abstract depiction of densification aspects. (a) During training a fraction of the weather stations are masked out from the input, while kept in the target. (b) To evaluate generalization to untrained locations, a set of weather stations represented by squares is never trained on and only used for evaluation. (c) To evaluate forecasts for the sparse locations for which data is available, these stations are fed as input during the evaluation as well. (d) The final forecasts uses the full set of training weather stations as input, and produces fully dense forecasts aided by spatial parameter sharing.}

\label{fig:abstract_densification}
\end{center}
\end{figure}

This paper presents MetNet-3, a weather forecasting neural network that is an advance over its predecessors
MetNet-1~\cite{sonderby2020metnet} and MetNet-2~\cite{Espeholt2022}.
Like its predecessors, MetNet-3 maintains the same high temporal prediction frequency of 2 minutes and spatial resolution of up to 1 km.
But MetNet-3 extends its lead time range from 12 hours to a full day range of 24 hours that involves dynamics well beyond extrapolation.
Besides rates of precipitation, that are  especially hard to predict due to their fast changing nature,
MetNet-3 also predicts another set of core weather variables including surface temperature, dew point, wind speed and direction.
While ground based radars provide dense precipitation measurements, observations that MetNet-3 uses for the other  variables come from just 942 points that correspond to weather stations spread out across Continental United Stated (CONUS). While NWP models transform the sparse points into a dense representation during data assimilation, MetNet-3 introduces a  process called \emph{densification} to achieve this that has four main aspects (see Figure~\ref{fig:abstract_densification}). The first aspect involves randomly dropping from the network's input a fraction of point observations during training, while keeping these observations as targets. The second and third aspects present two modes of evaluation of the densification, namely, evaluation on a hold-out set of stations that never appear during training to measure the network's ability to generalize spatially and perform implicit assimilation, and hyperlocal evaluation at just the specific points for which data is available. The last aspect is the inference step of densification, where the network relies on spatial parameter sharing to map all the sparse points given in the input to a fully dense image at the output, thereby producing dense forecasts.

Due to the challenge of incorporating all relevant sources of observational data that would provide a more complete picture of the recent conditions of the atmosphere,  MetNet-3, like MetNet-2, still relies on an assimilated NWP initial state that describes these conditions. This state includes a dense, albeit somewhat diverging, estimate of the surface variables that MetNet-3 predicts and can aid MetNet-3 in densifying its predictions into the future for these variables.

\section{Results}
\label{sec:results}

\begin{figure}[t]
    \newcommand{\vtitle}[1]{\rotatebox[origin=c]{90}{\hspace{-.3cm}#1}}
    \newcommand{\pred}[2]{
        \raisebox{-0.5\height}{%
            \begin{overpic}[width=.32\textwidth]{case_plots/patch_plots_hfmetar_windSpeed_None_538/#1.jpg}%
                \scriptsize
                \put(.7, 1){%
                    \textsf{\textbf{#2}}
                }
            \end{overpic}
        }%
    }
    \centering
    \setlength\tabcolsep{1.5pt}
    \begin{tabular}{@{}c c c c@{}}
        & ENS & MetNet-3 & OMO \\
        \vtitle{Mean} & \pred{ens_baseline_v4_360_mean}{} & \pred{metnet_180_mean}{} & \pred{passthrough_360_mean}{} \\
        & \multicolumn{3}{c}{\includegraphics[width=.4\textwidth]{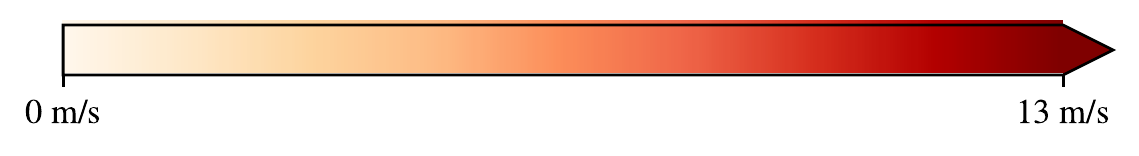}} \\
        \vtitle{Error} & \pred{ens_baseline_v4_360_diff}{MAE 3.36} & \pred{metnet_360_diff}{MAE 2.37} &  \\
        & \multicolumn{3}{c}{\includegraphics[width=.4\textwidth]{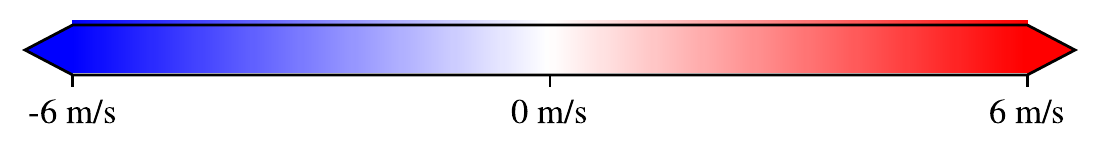}} \\
    \end{tabular}
    \caption{Case study for Sat Apr 23 2022 12:00 UTC featuring the Rocky Mountains of Colorado showing the mean of the ENS and MetNet-3 6~hour wind speed forecasts (top, left and center)
    along with the OMO stations ground truth (top, right) and the error of ENS and MetNet-3 on the individual weather stations (bottom).
    Circles and squares denote, respectively, training and test stations with
    MAEs calculated on both training and test stations.
    This example shows MetNet-3's ability to densify the targets, the higher spatial resolution of MetNet-3 as well as forecast precision on the weather stations.
    }
    \label{fig:example_wind}
\end{figure}

We evaluate MetNet-3 over CONUS on instantaneous rate of precipitation,
hourly accumulated precipitation, and the surface variables: 2m temperature, 2m dewpoint, 10m wind speed and 10m wind direction.

Ground truth estimates for instantaneous precipitation come from Multi-Radar/Multi-System (MRMS)~\cite{mrms} and rely on radar signals. 
The estimates have a high temporal frequency of 2 minutes and set the base lead time frequency of MetNet-3. On the other hand, the 1-hour accumulated precipitation estimates stem from both radar signals and ground rain gauges and have a temporal frequency of 60 minutes. MRMS is generally considered a high fidelity product~\cite{MultiRadarMultiSensorMRMSQuantitativePrecipitationEstimationInitialOperatingCapabilities} and following \cite{Espeholt2022} for evaluation we only use areas of MRMS where the radar fidelity is highest (see Supplement~C).

Ground truth observations for the surface variables come from the One Minute Observations (OMO) network of weather stations~\cite{metar} (see Supplement~C for a map of weather stations).
The weather stations include just 942 locations spread out across CONUS with observations stored for every 5th minute.
MetNet-3 applies densification to this network of weather stations.

\begin{figure}[t]
    \centering
    \begin{tabular}{@{} c c c c @{}}
        \textbf{1 h} & \textbf{6 h} & \textbf{12 h} & \textbf{24 h} \\
          \includegraphics[scale=.55]{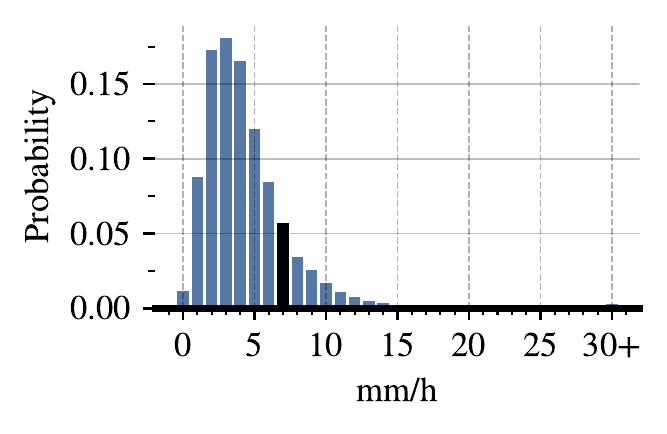} & \includegraphics[scale=.55]{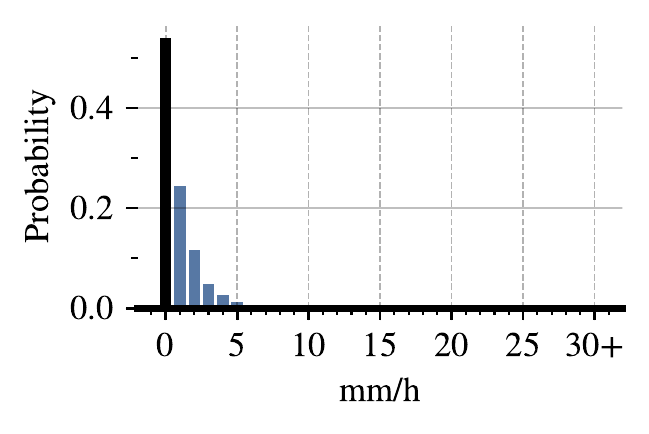} &
         \includegraphics[scale=.55]{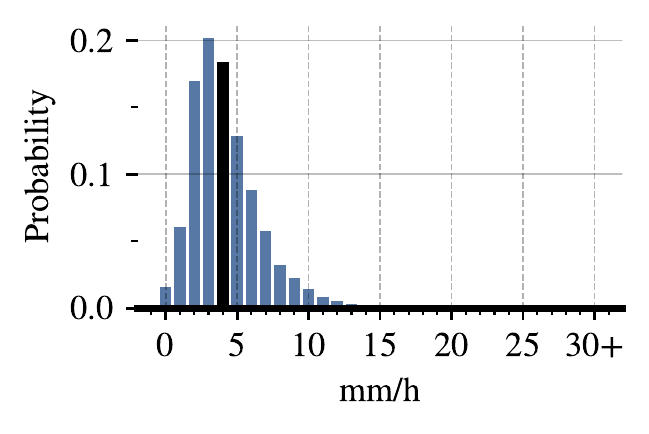}
         & \includegraphics[scale=.55]{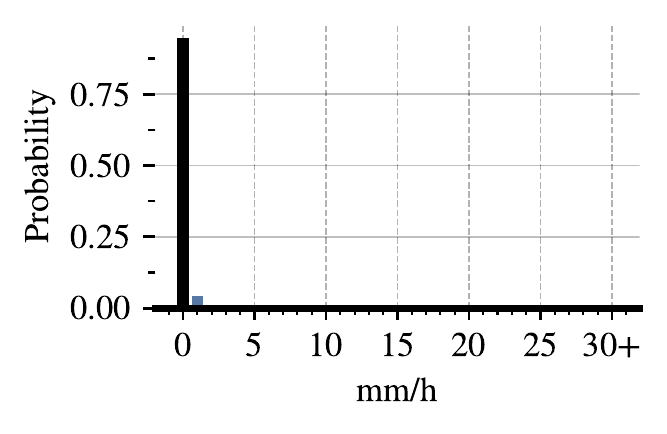}
    \end{tabular}
    \caption{An example of a precipitation rate distributions from MetNet-3 forecasts for a single location for different lead times. A black colored bar indicates MRMS precipitation ground truth rate.}
    \label{fig:distro}
\end{figure}

In contrast to NWPs that model uncertainty with ensemble forecasts,
MetNet-3 directly outputs a marginal probability distribution for each output variable and each location using a full categorical Softmax  that provides rich information beyond just the mean (see samples in Figure~\ref{fig:distro}).
We compare the probabilistic outputs of MetNet-3 with the outputs of advanced ensemble NWP models, including the
ensemble forecast (ENS) from the European Centre for Medium-Range Weather Forecasts (ECMWF) and the High Resolution Ensemble Forecast (HREF) from the National Oceanic and Atmospheric Administration of the US (NOAA). For reference, we also include single member forecasts from the High Resolution Rapid Refresh (HRRR) and the High Resolution Forecast (HRES) by NOAA and ECMWF, respectively.
We selected these models because they span the range of possible NWP models, as
the former two are ensembles, while the other two are single member NWP models, and two of them are global while the other two are designed for CONUS.
Figure~\ref{fig:baselines} summarizes basic characteristics of the baselines used in this work and of MetNet-3.
We compare the models' performance based on the metrics Continuous Ranked Probability Score (CRPS), Critical Success Index (CSI) and Mean Absolute Error (MAE). CRPS is particularly appropriate for comparison with ENS and HREF as they are ensembles of respectively 50 and 10 members and measures the accuracy of the full output distribution for all possible rates or amounts. This metric is one of the main metrics used for probabilistic forecasts and it plays an important role in guiding the development process for ENS at ECMWF~\cite{80865, ecmwf_crps_guiding}. More details on the evaluation protocol and the metrics used can be found in Supplement~D.

\begin{figure}[t]
\centering
\begin{tabular}{l l l p{2cm} p{2.5cm} p{2cm} l} 
 \toprule
 \textbf{Model} & \textbf{Region} & \textbf{Type} & \textbf{Operational Frequency} & \makecell[tl]{\textbf{Spatial} \\ \textbf{Resolution}} & \textbf{Temporal Resolution} & \\ %
 \midrule
 HRRR & CONUS & Deterministic & Hourly & 3 km & Hourly & \\
 HREF & CONUS  & Ensemble & Every 6 h & 3~km & Hourly \\
 HRES & Global & Deterministic & Every 6 h & 0.1\textdegree, 7--11~km & Hourly \\
 ENS & Global & Ensemble & Every 6 h & 0.2\textdegree, 14--22~km & Hourly \\
 \midrule %
 MetNet-3 & CONUS & Probabilistic & Every 10 min & 1-4 km & 2-5 min \\
 \bottomrule
\end{tabular}
\caption{Comparison of basic characteristics of physics-based baselines
used in this work and MetNet-3.
MetNet-3 forecasts precipitation at 1 km / 2 min resolution
and ground variables at 4 km / 5 min resolution.
MetNet-3 can be run more frequently than NWP models
because running the model is almost instant (about 1s for a single lead time)
and requires fewer computational resources than NWP models.
}
\label{fig:baselines}
\end{figure}

\subsection{Precipitation}

\begin{figure}[]
\centering
\begin{minipage}{.5\textwidth}
\centering
\textbf{(a) CRPS}\
\includegraphics[width=\textwidth]{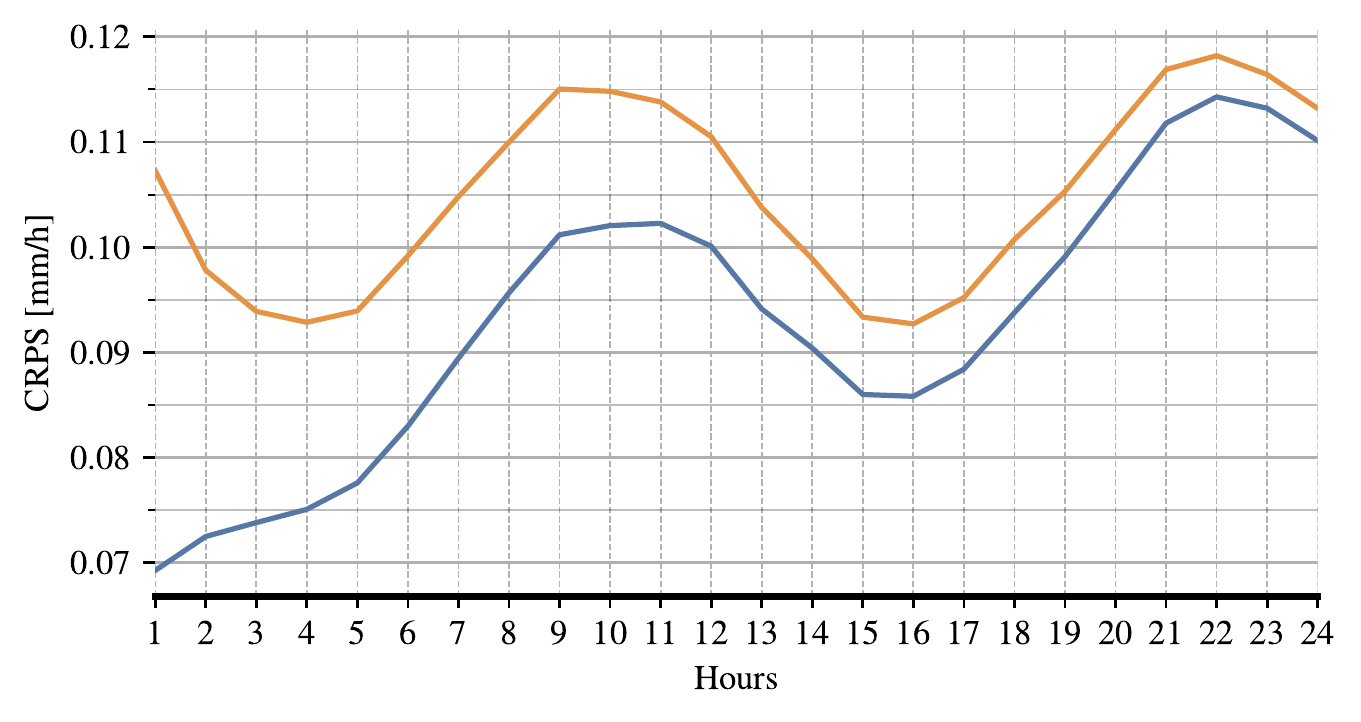}\
\end{minipage}%
\begin{minipage}{.5\textwidth}
\centering
\textbf{(b) CSI 1 mm/h}\
\includegraphics[width=\textwidth]{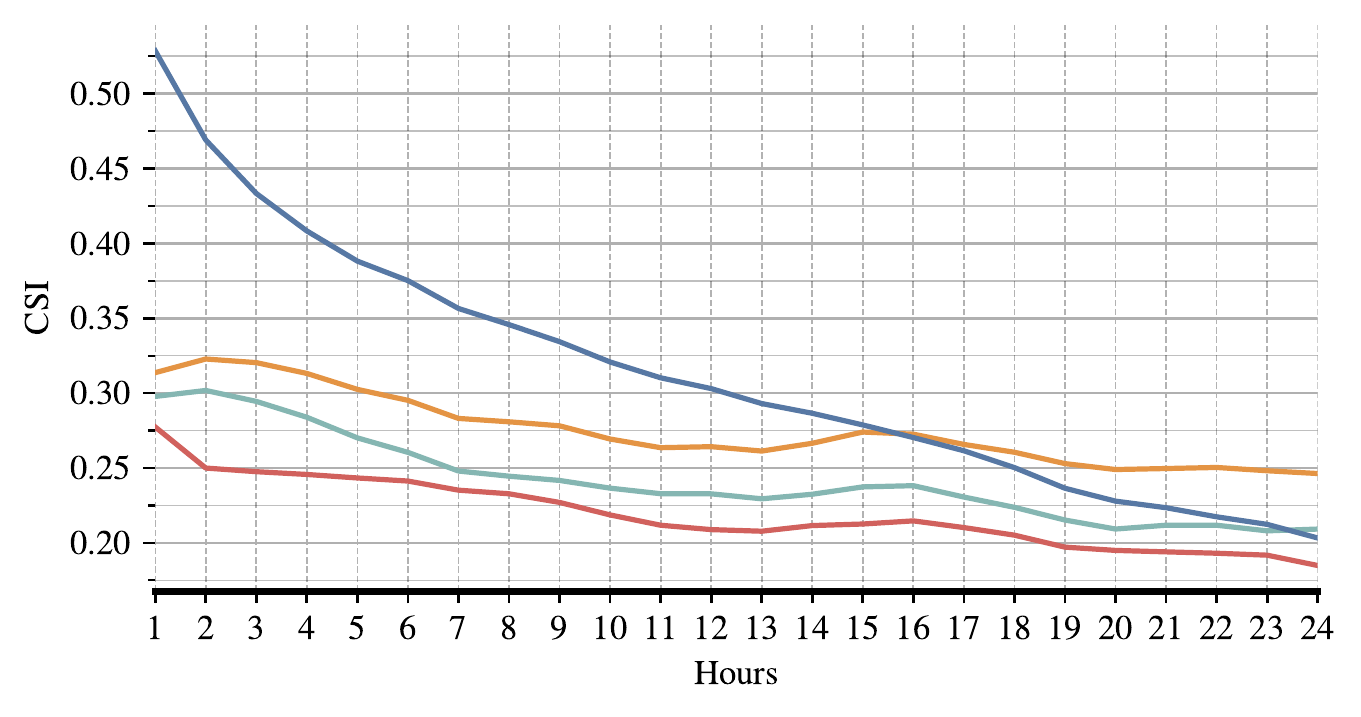}\
\end{minipage}\
\begin{minipage}{.5\textwidth}
\centering
\textbf{(c) CSI 4 mm/h}\
\includegraphics[width=\textwidth]{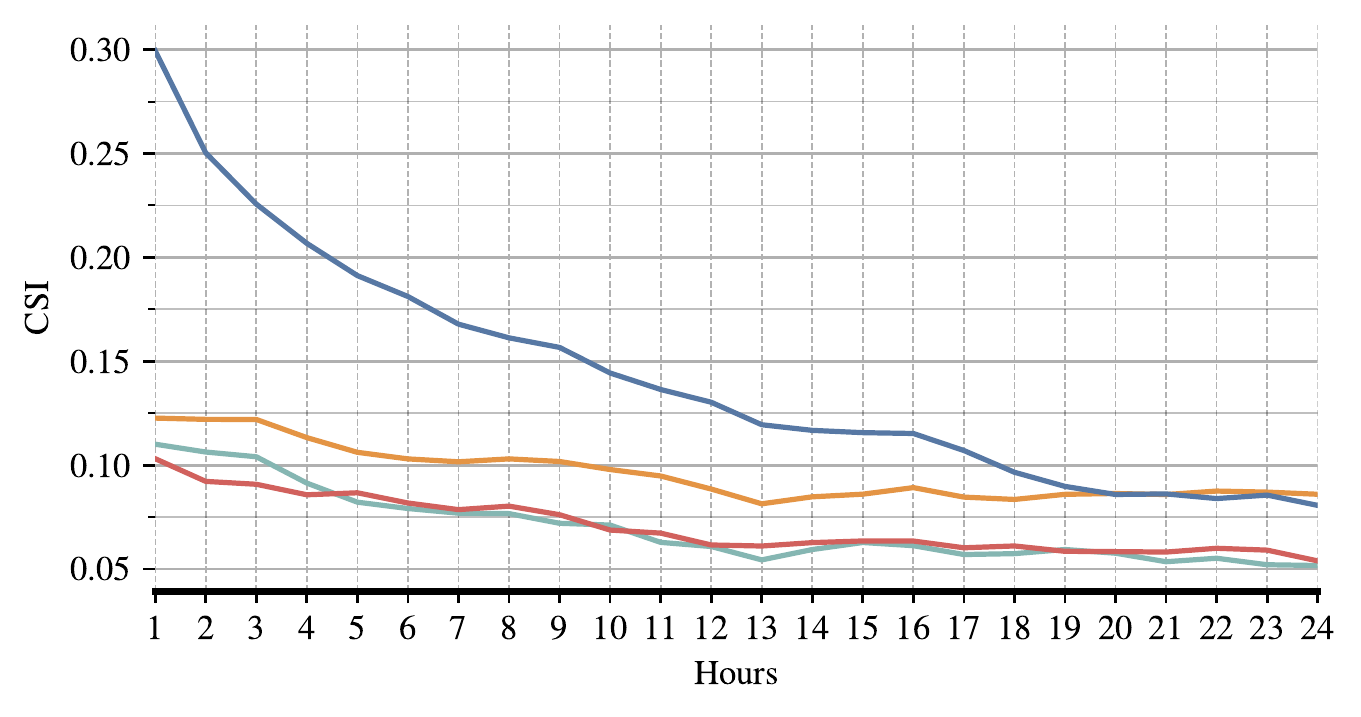}\
\end{minipage}%
\begin{minipage}{.5\textwidth}
\centering
\textbf{(d) CSI 8 mm/h}\
\includegraphics[width=\textwidth]{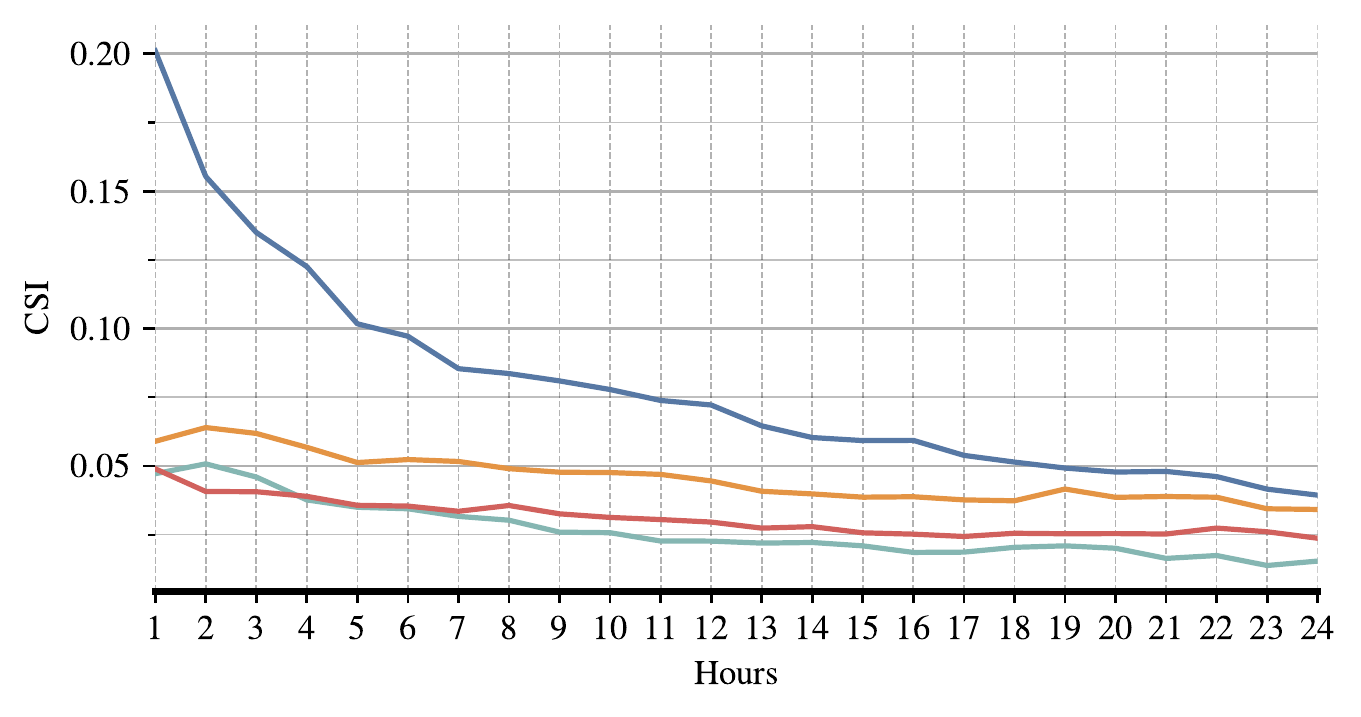}\
\end{minipage}\
\caption{Performance comparison between the probabilistic MetNet-3 and NWP baselines for instantaneous precipitation rate on 
CRPS (lower is better) and the categorical CSI (higher is better);
CRPS includes all precipitation rates, whereas the CSI plots are for light (1~mm/h), moderate (4~mm/h)
and heavy (8~mm/h) precipitation.
Deterministic baselines (HRRR and HRES) are ommited for clarity in the CRPS plot due to performing significantly worse than the probabilistic models.
Note, the thresholds for turning the probabilistic forecasts of MetNet-3 and ENS into deterministic forecasts for use in the CSI calculation, have been optimized on a validation set.
HREF is omitted in all plots due to instantaneous precipitation rate being unavailable.
See Supplement~E for the CSI plots for other rates, the CRPS plot with the deterministic baselines included
and an explanation for the CRPS lines being non-monotonic.
}
\label{fig:results_rate}

\includegraphics[height=5mm]{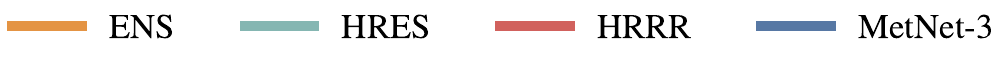}\
\end{figure}

 \begin{figure}[]
 \centering
 \begin{minipage}{.5\textwidth}
 \centering
 \textbf{(a) CRPS}\
 \includegraphics[width=\textwidth]{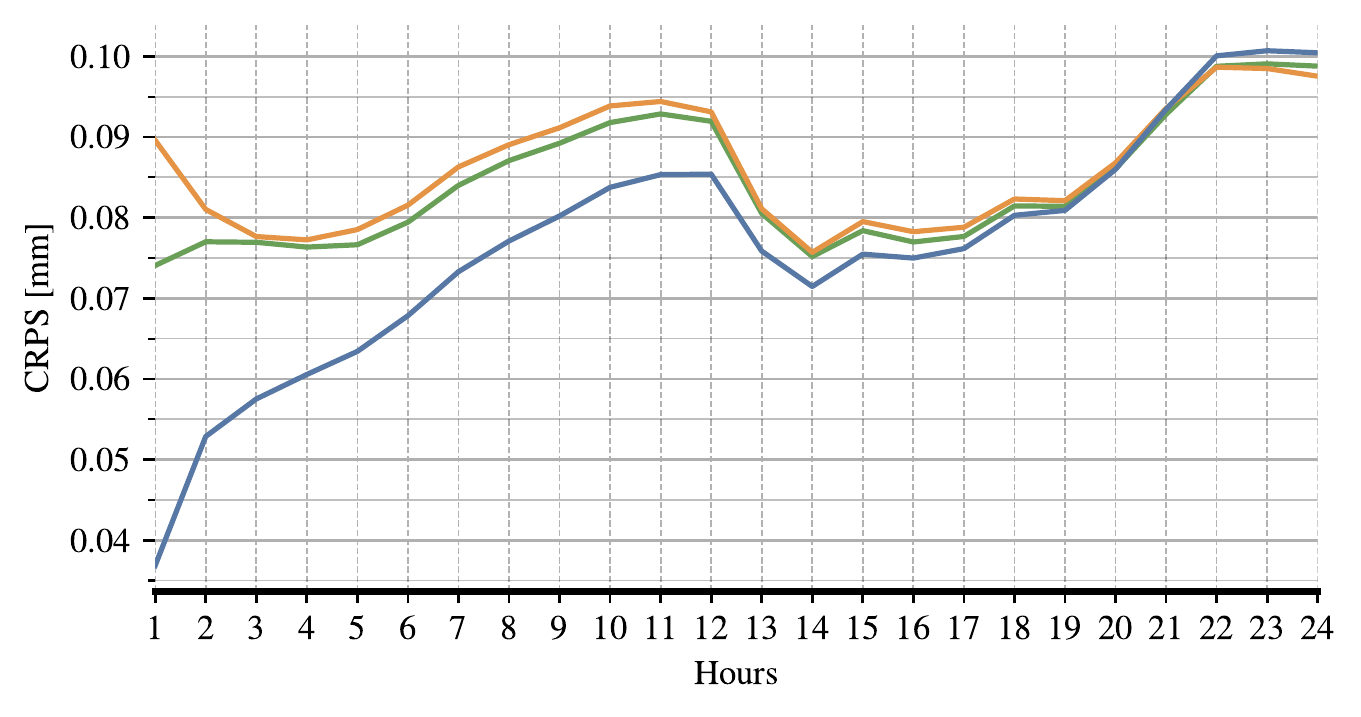}\
 \end{minipage}%
 \begin{minipage}{.5\textwidth}
 \centering
 \textbf{(b) CSI 4 mm}\
 \includegraphics[width=\textwidth]{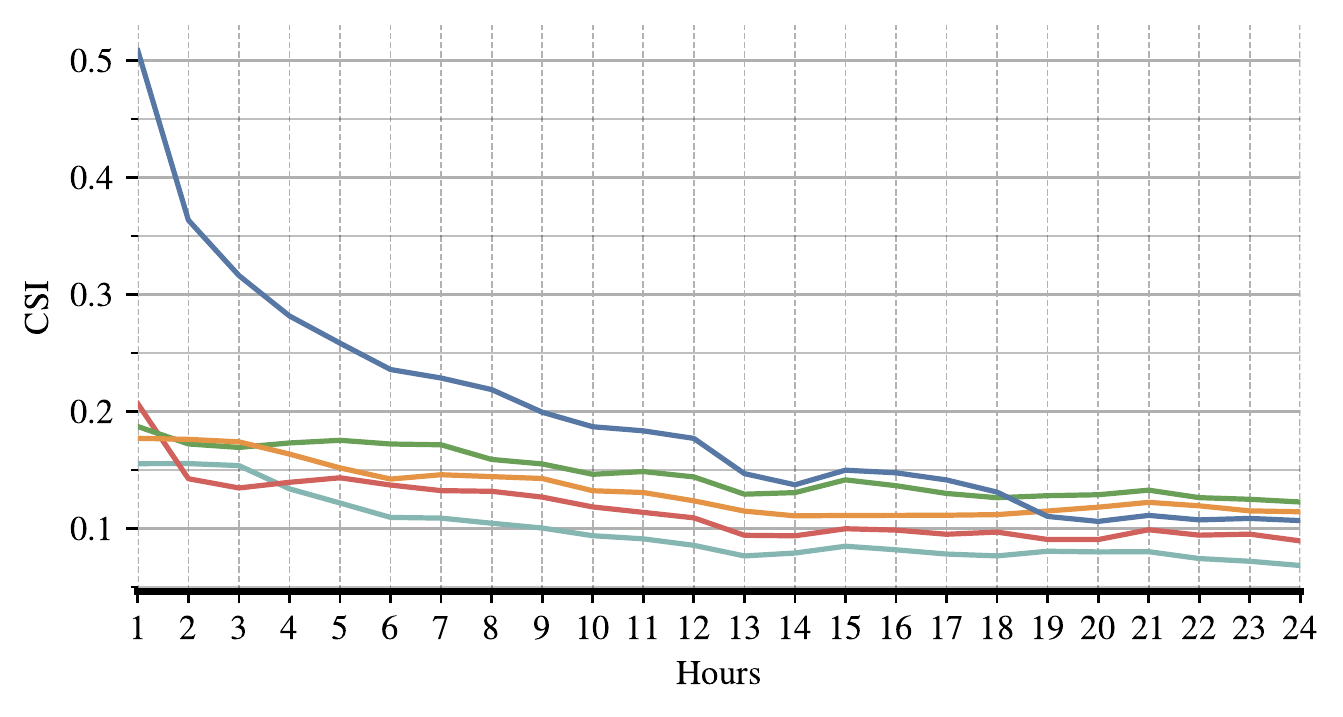}\
 \end{minipage}%
 \caption{Performance comparison between the probabilistic MetNet-3 and NWP baselines for hourly accumulated precipitation based on probabilistic
 CRPS (lower is better) and the categorical CSI (higher is better);
 CRPS includes all precipitation rates, whereas the CSI plot is for moderate (4 mm) precipitation.
 }
 \label{fig:results_cumulative}
 \includegraphics[height=5mm]{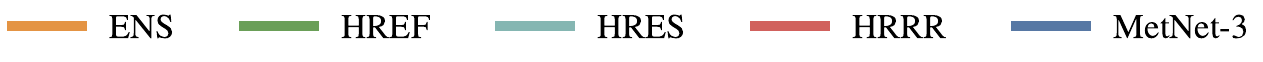}\
 \end{figure}

\begin{figure}
    \newcommand{\vtitle}[1]{\rotatebox[origin=c]{90}{\hspace{-.3cm}#1}}
    \newcommand{\pred}[2]{
        \raisebox{-0.5\height}{%
            \begin{overpic}[scale=.155]{\caseprefix#1.png}%
                \color{white}
                \scriptsize
                \put(.7, 1){%
                    \textsf{\textbf{#2}}
                }
            \end{overpic}
        }%
    }
    \newcommand{\caseprefix}{case_plots/paper_plots_probability_mrms_rate_1.0_1547683200/probability_mrms_rate_1.0_1547683200_}

    \centering
    \setlength\tabcolsep{1.5pt}
    \begin{tabular}{@{}c c c@{}}
    & \textbf{1 h} & \textbf{6 h} \\
    \vtitle{MRMS} & \pred{60_Passthrough_dark}{} & \pred{360_Passthrough_dark}{} \\[1.2cm]
    \vtitle{ENS} & \pred{60_ENS}{CSI .45} & \pred{360_ENS}{CSI .33} \\[1.2cm]
    \vtitle{MetNet-3} & \pred{60_Multi-21}{CSI .59} & \pred{360_Multi-21}{CSI.41} \\[1.2cm]
    & \textbf{12 h} & \textbf{24 h} \\
    \vtitle{MRMS} & \pred{720_Passthrough_dark}{} & \pred{1440_Passthrough_dark}{} \\[1.2cm]
    \vtitle{ENS} & \pred{720_ENS}{CSI .26} & \pred{1440_ENS}{CSI .18} \\[1.2cm]
    \vtitle{MetNet-3} & \pred{720_Multi-21}{CSI .28} & \pred{1440_Multi-21}{CSI .17} \\[1.2cm]
    \end{tabular}
    \includegraphics[width=.5\textwidth]{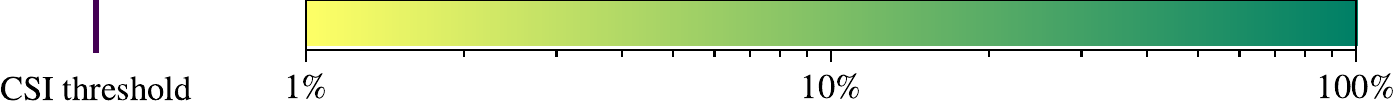}
    \caption{Case study for Thu Jan 17 2019 00:00 UTC showing the probability of instantaneous precipitation rate being above 1~mm/h on CONUS.
    The maps also show the prediction threshold when optimized towards CSI (dark blue contours) as well as the CSI values (lower left corners) calculated on the evaluation mask (Figure~2 in Supplement~C).
    This specific case study shows the formation of a new large precipitation pattern in central US and \emph{not} just extrapolation of existing patterns.}
    
    \label{fig:case_study_precipitation}
\end{figure}

The first main result is that MetNet-3 obtains a higher CRPS than ENS
for forecasting the rate of instantaneous precipitation over the whole lead time range of 24~hours suggesting that averaged across all rates MetNet-3's performance is superior to that of ENS (Figure~\ref{fig:results_rate}a).
When thresholding the MetNet-3 and ENS output probabilities, optimized on the categorical metric CSI, MetNet-3 outperforms ENS for the first 15 hours of lead time for light (1~mm/h) precipitation (Figure~\ref{fig:results_rate}b) and outperforms ENS on the whole lead time range of 24~hours for heavy (8~mm/h) precipitation (Figure~\ref{fig:results_rate}d).
The skill gap between MetNet-3 and ENS is greatest in relative terms at the earliest hours
and decreases gradually over time.
In Figure~\ref{fig:case_study_precipitation}, we show a 24~hour forecast of CONUS demonstrating the spread of MetNet-3 and ENS probability distributions and the ability of MetNet-3 to predict new precipitation formations. 

Hourly accumulated precipitation estimates stemming from MRMS have a frequency of only 60 minutes and provide substantially fewer distinct data frames over the same period of time that can be used for training MetNet-3 relative to those for instantaneous precipitation. Nevertheless,  we find that MetNet-3 outperforms both of the multi-member NWP baselines ENS and HREF on the CRPS metric up to the first 19~hours of lead time
(Figure~\ref{fig:results_cumulative}a).
When thresholding based on the CSI metric, MetNet-3 outperforms the baselines for the first 18 hours of lead time for moderate (4~mm/h) precipitation (Figure~\ref{fig:results_cumulative}b).
Detailed results can be found in the Supplement~E. With these results MetNet-3 extends the lead time advantage over probabilistic NWPs for observation-based models from the 12 hours achieved by MetNet-2 all the way to 19 hours.

\subsection{Sparse Surface Variables}

Learning from weather station observations is challenging because the OMO ground truth targets
are only available at 942 weather stations throughout CONUS, while weather models are required to predict weather variables at all locations.
Training and evaluating on the same weather stations could lead to a situation where a model
performs well on the locations that  it was trained on but poorly elsewhere.
To make sure that MetNet-3 performs well throughout CONUS, we apply the densification procedure for these sparse variables and we perform evaluation
on a hold-out set of 20\% randomly selected weather stations that are not used during MetNet-3 training and are not fed as input during evaluation either. In Figure~\ref{fig:example_wind}, we show a case study of MetNet-3's ability to predict a densified forecast of the surface variables as well as errors on training and test weather stations. %

\begin{figure}[h!]
\centering
\begin{minipage}{.5\textwidth}
\centering
\textbf{(a) Temperature CRPS}\
\includegraphics[width=\textwidth]{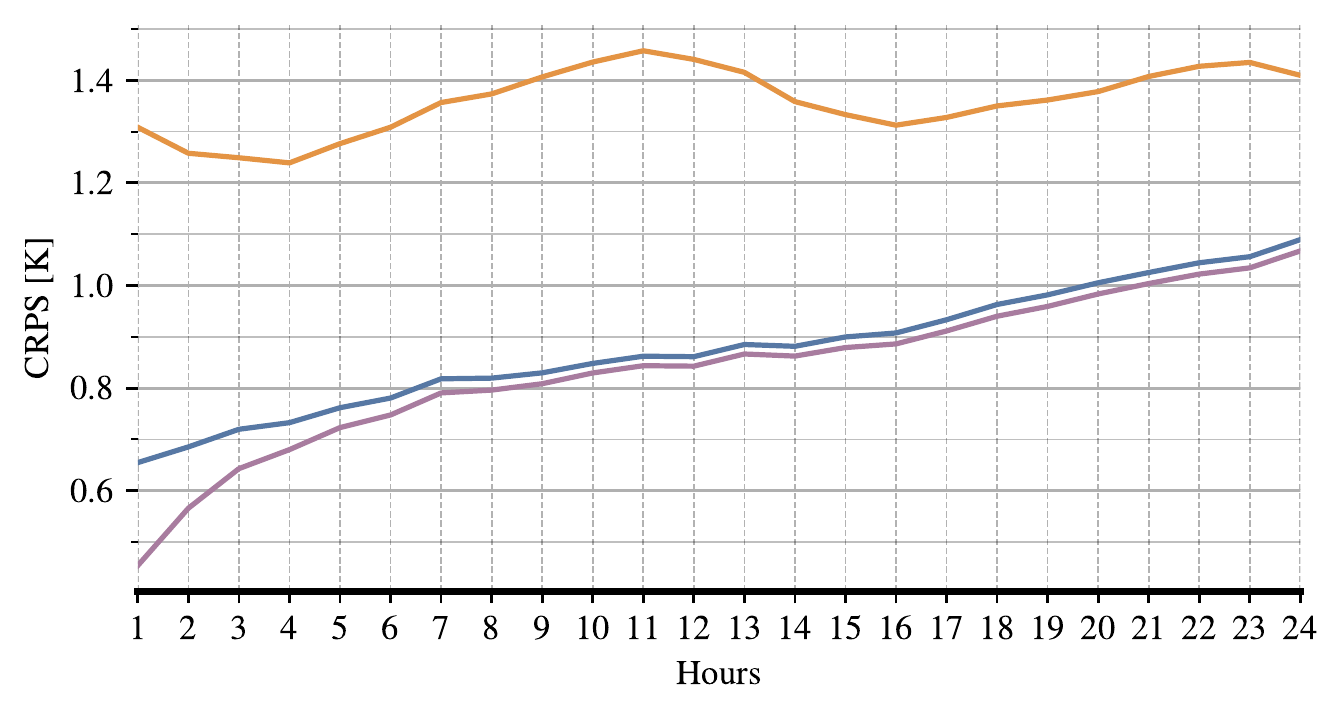}\
\end{minipage}%
\begin{minipage}{.5\textwidth}
\centering
\textbf{(b) Temperature MAE}\
\includegraphics[width=\textwidth]{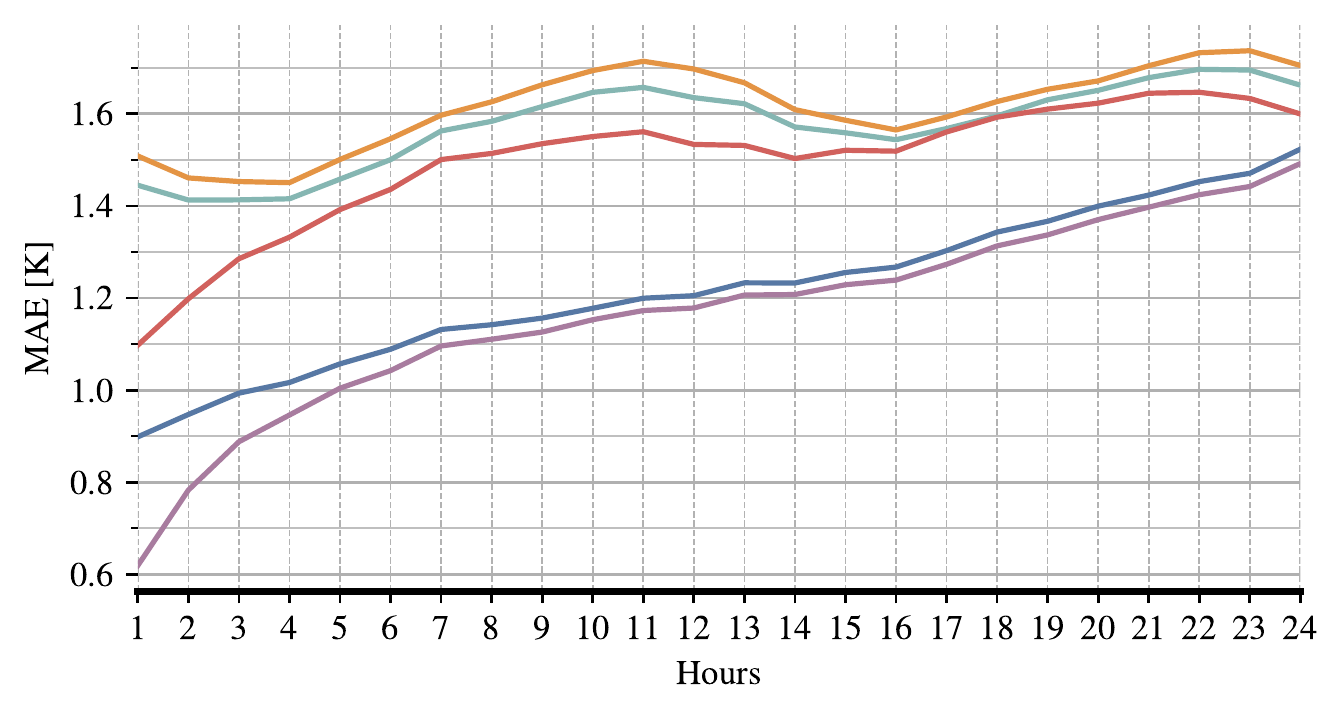}\
\end{minipage}\
\begin{minipage}{.5\textwidth}
\centering
\textbf{(c) Dew point CRPS}\
\includegraphics[width=\textwidth]{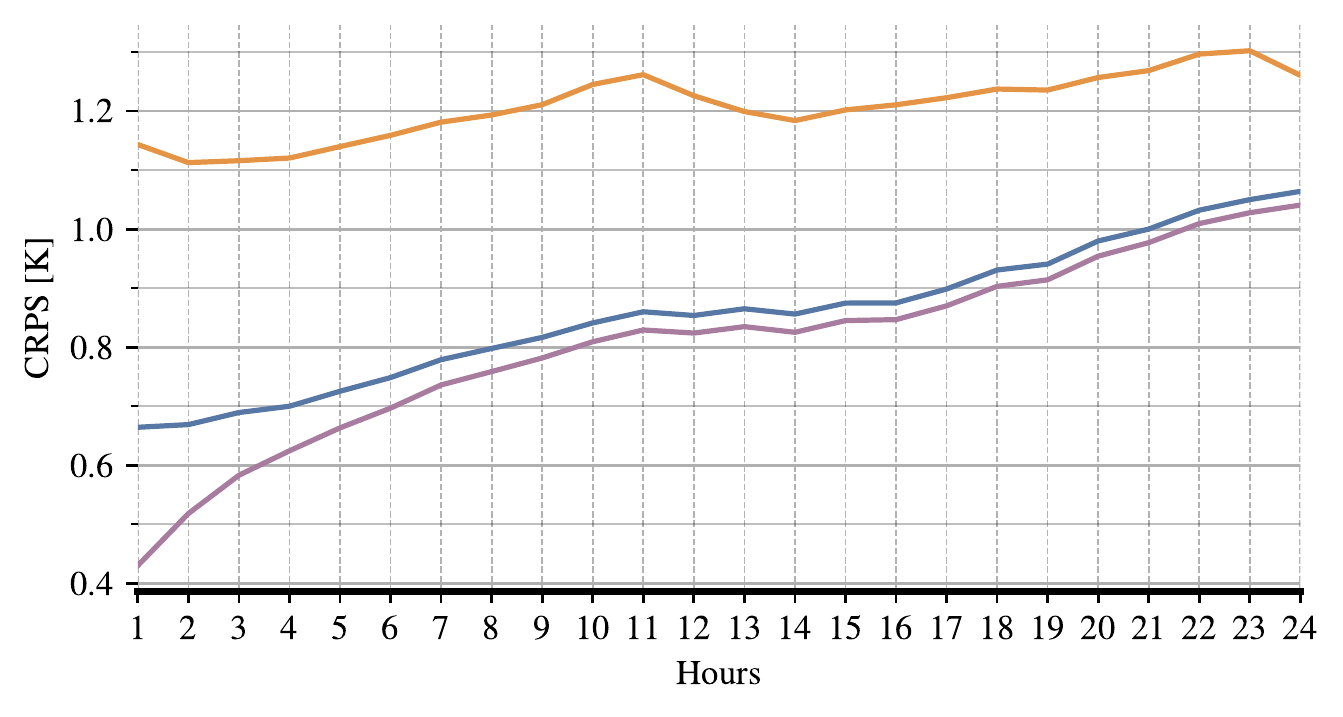}\
\end{minipage}%
\begin{minipage}{.5\textwidth}
\centering
\textbf{(d) Dew point MAE}\
\includegraphics[width=\textwidth]{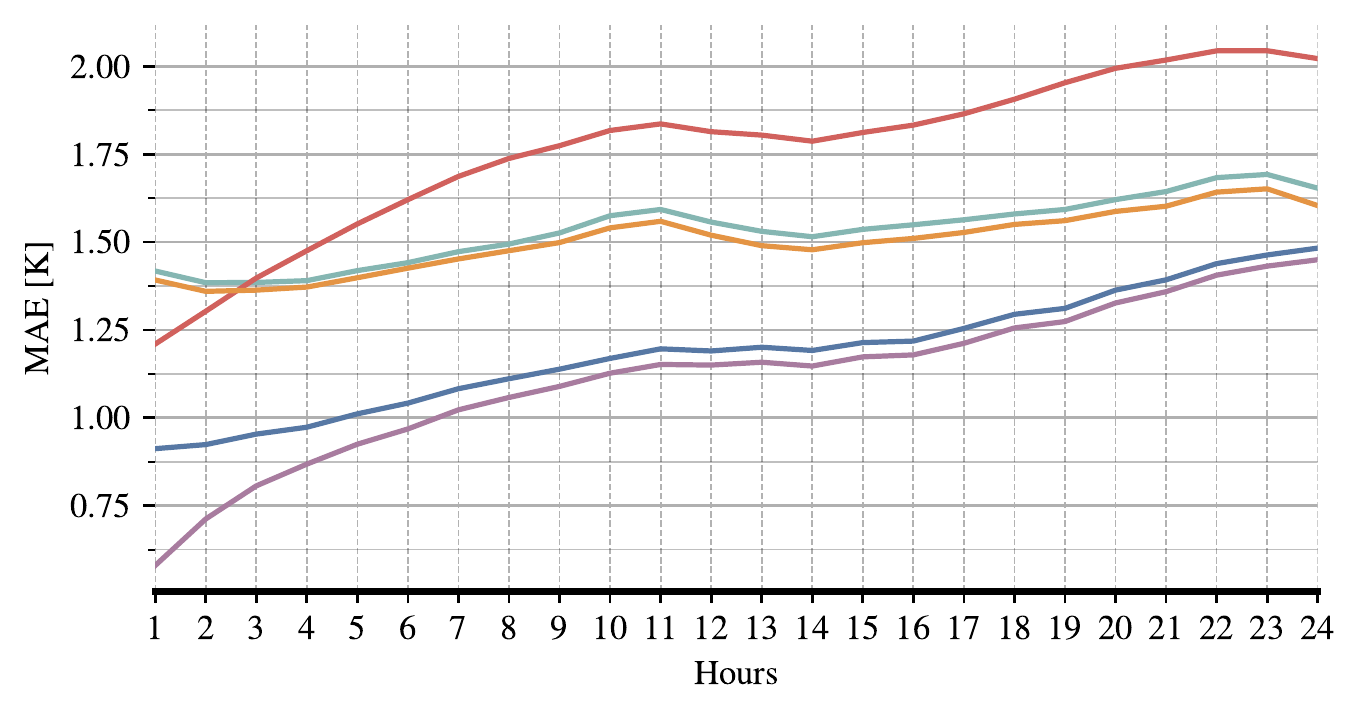}\
\end{minipage}\
\begin{minipage}{.5\textwidth}
\centering
\textbf{(e) Wind speed CRPS}\
\includegraphics[width=\textwidth]{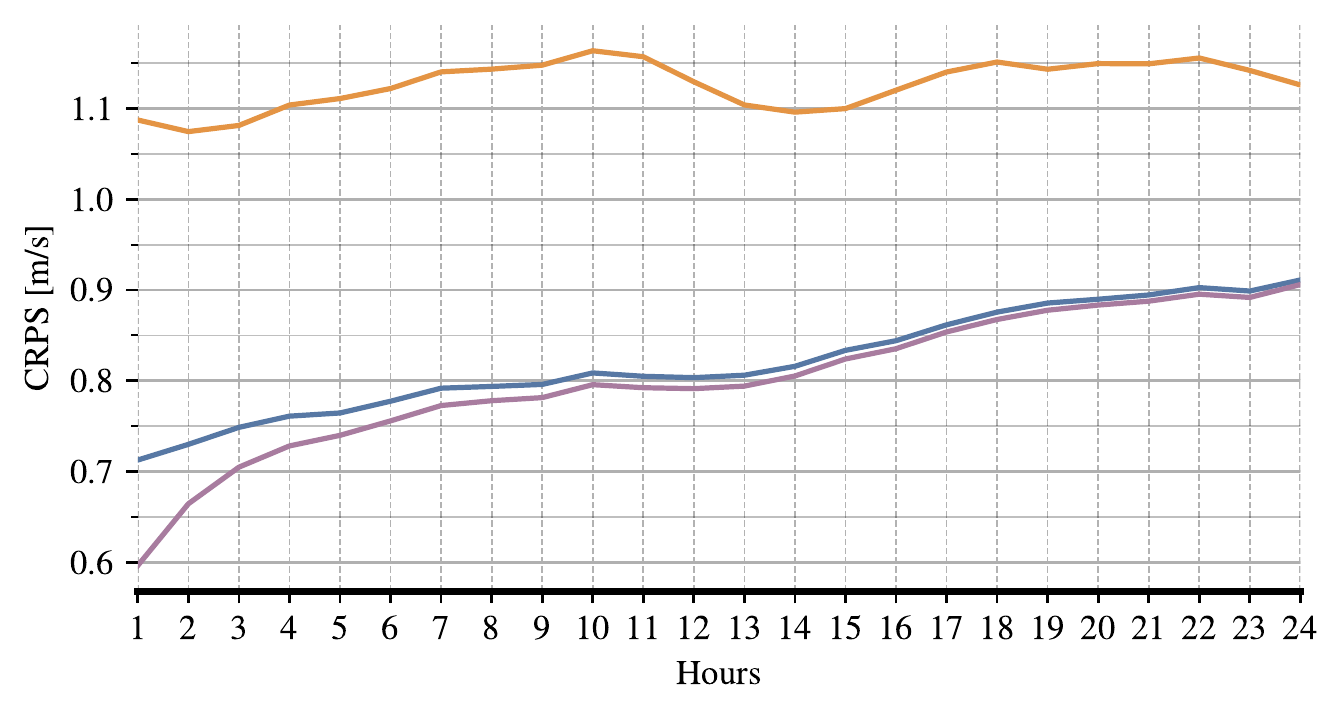}\
\end{minipage}%
\begin{minipage}{.5\textwidth}
\centering
\textbf{(f) Wind speed MAE}\
\includegraphics[width=\textwidth]{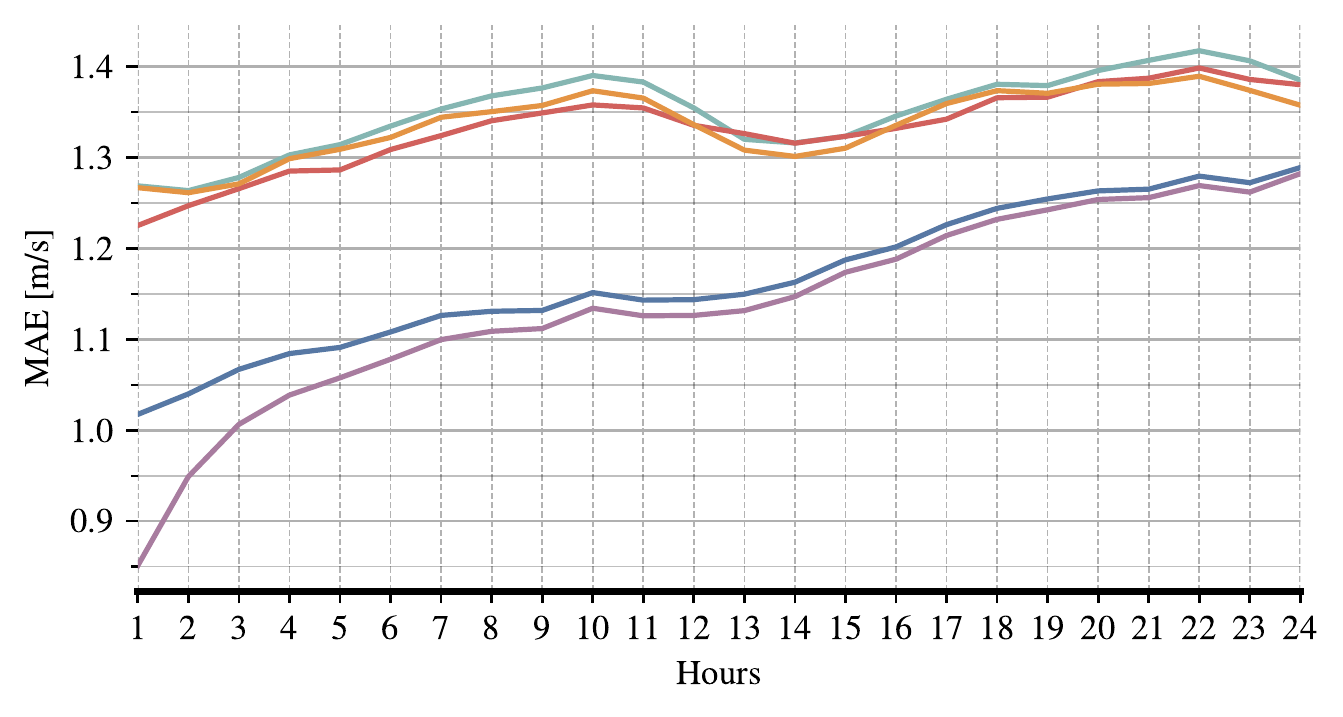}\
\end{minipage}\
\caption{Performance comparison between the probabilistic MetNet-3 and NWP baselines for ground variables: temperature, dew point and wind speed
based on CRPS and MAE (lower is better).
Deterministic baselines (HRRR and HRES) are omitted in the CRPS plots because CRPS take the full forecast distribution into account
and is therefore more appropriate for probabilistic models.
Results for wind components can be found in the Supplement~E. For these variables, we did not have HREF variables available.
}
\label{fig:results_metar}
\includegraphics[height=5mm]{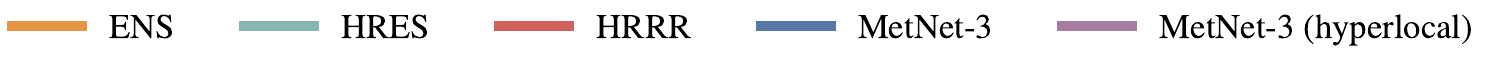}\
\end{figure}

\begin{figure}[h]
    \centering
    \begin{minipage}{1\textwidth}
    \centering
    \textbf{(a) Temperature}\
    \includegraphics[width=\textwidth]{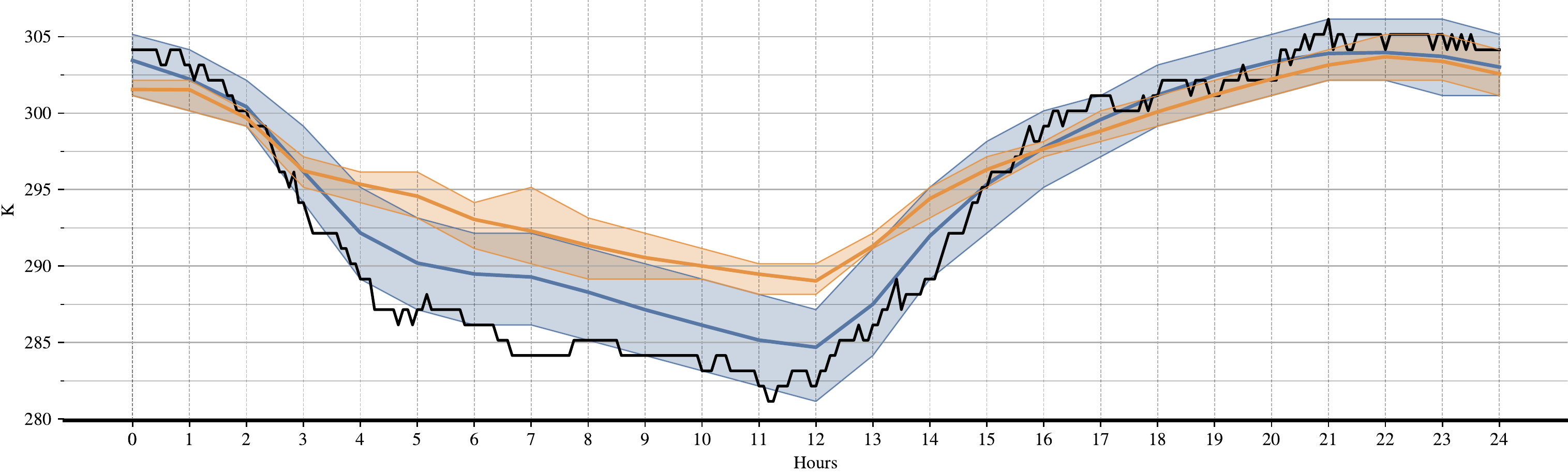}\
    \end{minipage}\\%
    \begin{minipage}{1\textwidth}
    \centering
    \textbf{(b) Wind speed}\
    \includegraphics[width=\textwidth]{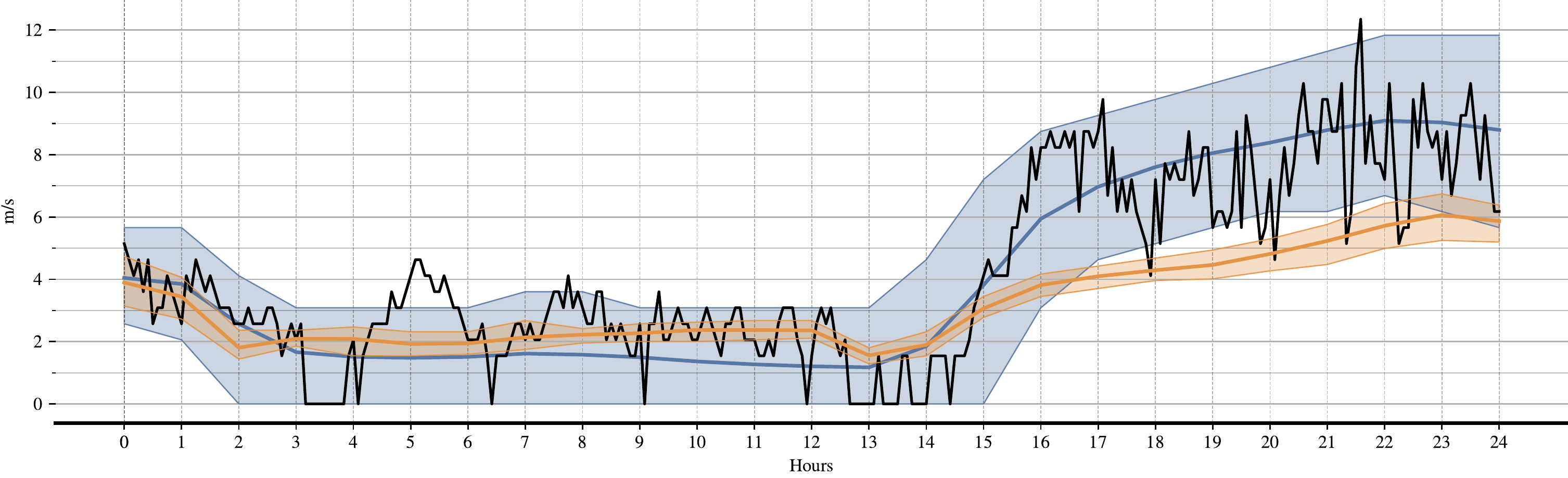}\
    \includegraphics[scale=.7]{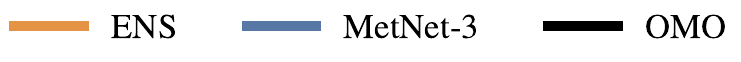}
    \end{minipage}\
    \caption{Case study for Thu Jun 10 2021 00:00 UTC
    comparing a MetNet-3 forecast and an ENS forecast for a single location (117.22\textdegree W, 33.91\textdegree N): 
    Bold lines depict the means of the forecast distributions,
    and shaded areas correspond to
    80\% confidence interval based on the 10th- and 90th-quantile of the forecasted
    distribution in the case of MetNet-3 and the ensemble distribution in the case of ENS.
    \label{fig:example_metar_nonhyperlocal}
    }
\end{figure}

Like for precipitation, the surface  values are discretized into bins and a Softmax layer with a categorical loss is used to predict them. MetNet-3 obtains better CRPS and MAE for all surface variables than multi-member ENS  for the full
24~hour range of lead times (Figure~\ref{fig:results_metar}).
In terms of CRPS that takes the full forecast distribution into account MetNet-3 shows a significant gain.
For all surface variables, MetNet-3 CRPS values at all lead times up to 24 hours are better than ENS's highest CRPS, which is obtained at the shortest lead time of 1 hour ahead.
This also suggests that ensemble NWPs do not model the forecast distribution particularly well in this time range. Figure~\ref{fig:example_metar_nonhyperlocal} shows examples of MetNet-3 forecast at a single location for temperature and wind speed.
Most of the observed values fall into the 80\% confidence interval of MetNet-3's predicted distribution, whereas ENS's predicted distribution is very peaked and under-dispersed.

The results in terms of the MAE metric depict a similar picture where MetNet-3 achieves much better results than both the multi-member and the single-member baselines, e.g.
the 20~hour MetNet-3 temperature forecast has similar MAE as a 5~hour forecast from the best performing baseline.
In the hyperlocal setting, where the values of the test weather stations are given as input to the network during evaluation, the results improve further especially in the early lead times.

\section{Discussion}\label{sec:discussion}

\begin{figure}[h]
\centering
\begin{minipage}{.5\textwidth}
    \centering
    \includegraphics[width=\textwidth]{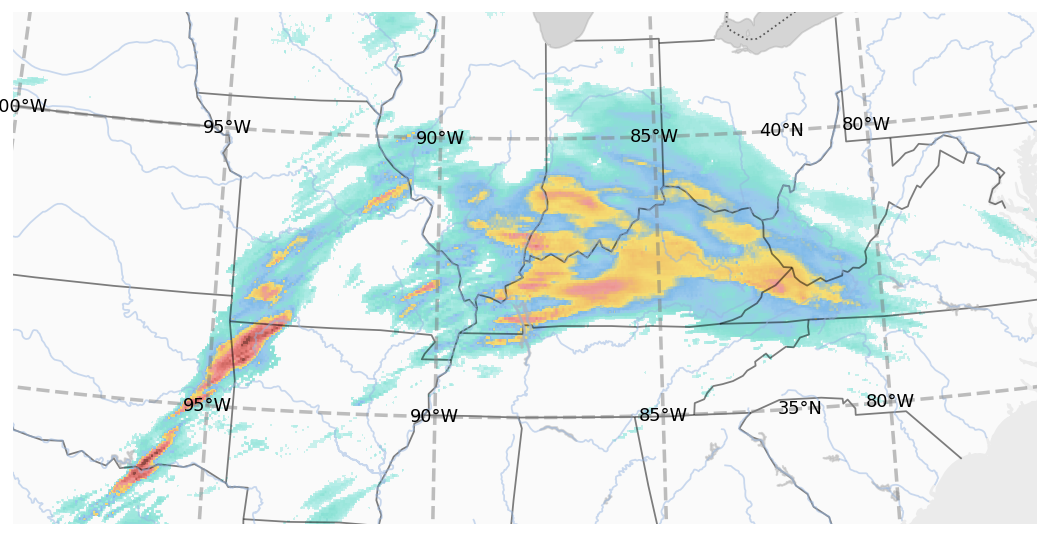}\\
     \end{minipage}%
    \begin{minipage}{.5\textwidth}   
    \centering
    \includegraphics[width=\textwidth]{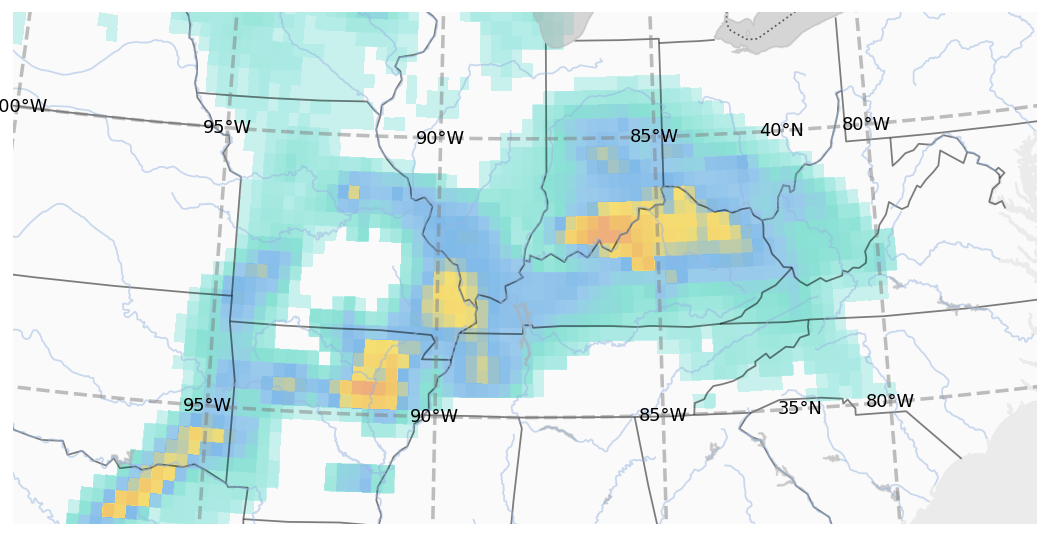}\\
    \end{minipage}\\
    \includegraphics[width=.3\textwidth]{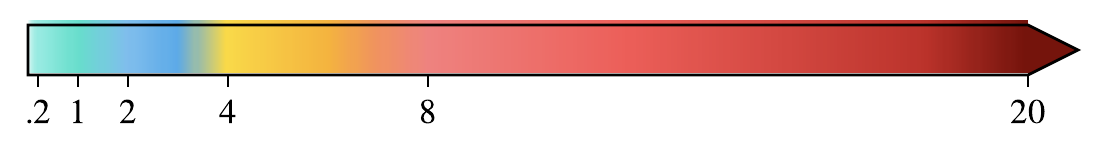}
    \caption{Hourly accumulated precipitation in millimeter accordingly to
    gauge-corrected MRMS product (Left) and ERA5 reanalysis data (Right), for the timestamp Sat Nov 30 2019 12:00 UTC.
    }
    \label{fig:assimilation_era5_vs_mrms}
\end{figure}

MetNet-3 considers observational data as the highest fidelity target and as the ground truth for evaluation and training. An alternative would be to use assimilation or reanalysis states from NWP models as targets where the reanalysis state, as opposed to assimilation state, integrates not just past and present information about the atmosphere, but also future information~\cite{weatherbench, keisler2022forecasting, fourcastnet1, fourcastnet2, panguweather, graphcast, climax, fengwu}. However, this option has limitations. First, we find a significant mismatch between ground truth observations and the values of the same variables provided by the NWP states. Figure~\ref{fig:assimilation_era5_vs_mrms} gives an example of the core variable of hourly accumulated precipitation in the NWP reanalysis state from ERA5 \cite{era5} and of the corresponding estimates from the gauge-corrected MRMS; the comparison shows  a marked mismatch between the two. Similarly, for surface temperature, we obtain an MAE of 1.5~C and 0.9~C between OMO weather stations and the HRES and HRRR assimilation states, respectively, a margin of error that is at times larger than that of MetNet-3's forecasts themselves. See Supplement~A for an example of mismatch with HRES on instantaneous precipitation. Evaluating or training against such states makes it hard to gauge the accuracy of the resulting model.
A second limitation of reanalysis or assimilation states is that they are spatially and temporally coarse. ERA5 for example has a spatial resolution of approximately 25~km and a frequency of 6~hours. A strength of neural models is that their computation only grows linearly with temporal frequency and spatial resolution and using coarse targets, like ERA5, limits the potential of what neural models can learn. In addition, densification from point data allows the neural model to choose an arbitrarily coarse output resolution by assigning the point data to the respective grid cell. Yet another limitation concerns the real world latency of such targets. A lag of 6 hours like that of the ENS model has a large impact on short term performance, as can be seen in Section~\ref{sec:results}.
On top of that, ENS only runs 4 times per day and thus provides forecasts that rely on stale atmospheric information from up to 12 hours prior to the forecast time. This has a large effect on the operational performance of a model and for this reason MetNet-3 relies on observations with latency on the order of minutes and on the HRRR state whose latency is 55~minutes. MetNet-3 then takes another 10 minutes to generate a forecast for all of CONUS for all lead times every two minutes up to 24 hours in advance. If adjusted for operational latency, MetNet-3's gains over the NWP baselines would be larger than reported in Section~\ref{sec:results}.

When compared to MetNet-2, MetNet-3 shows a leap forward in performance.
Figure~11 in Supplement~E shows how the multiple innovations of MetNet-3 lead to a substantial gain.
MetNet-2 in turn obtained a similar improvement over the original MetNet.
This paints a picture where neural weather models keep on improving due to better architectures and observation sources. MetNet-3 still uses a tiny fraction of all available atmospheric data.

\section{Methods}

\subsection{Dataset Creation}

The data for MetNet-3 comes in input-output pairs where the inputs include radar (estimated precipitation rate and type)
data from the last 90 mins,
sparse OMO weather station reports from the last 6~hours,
images from GOES satellites,
assimilated weather state, latitude and longitude information, altitude information and current time, and outputs correspond to the future radar precipitation estimates (instantaneous radar-only precipitation rates as well as gauge-corrected hourly accumulations), measurements from ground weather stations (temperature, dew point, pressure and wind speed and direction) and assimilated weather state (the latter is only used to improve the model training and we do not treat it as ground truth).
See Table~\ref{tab:inputs} for more information on the inputs used, and Table~\ref{tab:outputs} for more information on the targets used.
The available data spans a period from July 2017 to September 2022. The training, validation and test data sets are generated without overlap from periods in sequence. Successive periods of 19 days training data, 1 day blackout, 2 day validation data, 1 day blackout, 2.5 days test data and 1 day blackout are used to sample, respectively, training, validation and test data with no sampling in the blackout periods.
To increase the number of training samples, we temporarily interpolate targets in the train split using linear interpolation whenever
the observation for the exact lead time is not available.
Spatially, the target patches are sampled randomly from intersections on a grid over the CONUS region spaced at .5 degrees in longitude and latitude.

For surface variables,
we take the OMO station point measurements and map them to a 4~km by 4~km pixel
in which the station lies.
If there are multiple stations in a given region, we take the average of their
measurements. For all 942 weather stations, only 12 pairs of stations are within a distance for which it is necessary to average the weather station variables.
Apart from temporal splits, we also divide OMO stations into two groups: 757 training
stations and 185 test stations (Supplement~C, Figure~3).
The data from test stations is not used in any way during training and we only report the results
on the test stations.
Moreover, we normally do not include past observations from the tests stations in MetNet-3 inputs even during evaluation, so
that the model  does not have any information about test station exact locations and produces ground forecasts representative
of the full 4 km by 4 km output squares.
Including past observations from the test stations allows MetNet-3 to bespoke the forecast to a particular weather station
and results in hyperlocal forecasts.

\subsection{Model and Architecture}

On a high level, MetNet-3 neural network consists of three parts:
topographical embeddings, U-Net~\citep{unet} backbone and a MaxVit~\citep{maxvit} transformer for capturing long-range interactions.
The whole network has 227M trainable parameters.

\subsubsection{Topographical Embeddings}
It is common to feed neural weather models multiple time-independent variables containing topographical information
like sea-land mask \citep{Espeholt2022}. Instead of manually selecting and preparing this kind of information,
we use a novel technique of \emph{topographical embeddings}, which allows the network to automatically
discover relevant topographical information and store it in embedding. More precisely, we allocate
a grid of embeddings with a stride of 4~km where each point is associated with 20 scalar parameters.
For each input example, we calculate the topographical embedding of each input pixel center
by bilinearly interpolating the embeddings from the grid.
The embedding parameters are trained together with other model parameters
similarly to embeddings used in NLP.

\subsubsection{Network Architecture}

 \begin{table}[t]
\begin{center}
\begin{tabular}{ l c  c  c c} 
\toprule
 \textbf{Input} & \textbf{Context size} & \textbf{Resolution} & \textbf{\#Channels} & \textbf{\#Time Slices} \\ \midrule
    {Radar MRMS} & 2496 km & 4 km & 2 & 11  \\
    {Weather stations OMO} & 2496 km & 4 km & 14 & 9 \\
    {Elevation} & 2496 km & 4 km & 1 & 1 \\
    {Geographical coordinates} & 2496 km & 4 km & 2 & 1  \\
    {Topographical embeddings} & 2496 km & 4 km & 20 & 1  \\
    {HRRR assimilation} &  2496 km & 4 km & 617+1 & 1  \\
    \midrule
    {Low-resolution Radar MRMS} & 4992 km & 8 km & 1 & 1  \\
    {GOES Satellites} & 4992 km & 8 km & 16 & 1  \\
    \bottomrule
\end{tabular}
\caption{MetNet-3 spatial inputs.
For HRRR assimilation, the model is given 617 channels from the assimilated state as well as one channel
containing information about how stale the HRRR state is.
Apart from the spatial inputs, MetNet-3 is also given the time when the prediction is made (month, day, hour and minute)
and the lead time.
}
\label{tab:inputs}
\end{center}
\end{table}

\begin{table}[h]
\begin{center}
\begin{tabular}{ l l l c l l} 
\toprule
 \textbf{Target} & \textbf{Source} & \textbf{Resolution} & \textbf{\#Channels} & \textbf{Output Type} & \textbf{Loss Function} \\ \midrule
    Precipitation & MRMS & 1 km / 2 min & 2 & Categorical & Cross Entropy \\
    Surface Variables & OMO & 4 km / 5 min & 6  & Categorical & Cross Entropy \\
    Assimilation & HRRR & 4 km / 1 h & 617  & Deterministic & Mean Squared Error \\
    \bottomrule
\end{tabular}
\caption{MetNet-3 targets.
For precipitation, we use radar-only instanteneous precipitation estimates from MRMS as well as
hourly precipitation accumulations which also take rain gauges into account.
For surface variables, we use temperature, dew point and wind (speed, direction and 2 components) as
reported from OMO.
}
\label{tab:outputs}
\end{center}
\end{table}

The network architecture is presented in Figure~\ref{fig:models_flow}.
The network uses two types of inputs: high-resolution, small-context (2496 km by 2496 km at 4 km resolution) ones
and low-resolution, large-context ones (4992 km by 4992 km at 8 km resolution).
All time slices from different high-resolution inputs (see Table~\ref{tab:inputs})
are first concatenated across the channel dimension, then current time is also concatenated across the channel dimension, which
results in an 624 x 624 x 793 input image.

Data is then processed by a U-Net backbone, which starts with
applying two convolutional ResNet blocks~\citep{resnet} and downsampling the data to 8 km resolution.
We then pad the internal representation spatially with zeros to 4992 km by 4992 km square and concatenate with the
low-resolution, large-context inputs.
Afterward, we again apply two convolutional ResNet blocks and downsample the representation to 16 km resolution.
Convolutional ResNet blocks can only handle local interactions and
for longer lead times close to 24~hours, the targets may depend on the entire input.
In order to facilitate that, we process the data at 16 km resolution
using a modified version of MaxVit~\citep{maxvit} network.
MaxVit is a version of Vision Transformer~(ViT,~\citep{vit}) with attention
over local neighbourhood as well as global gridded attention.
We modify the MaxVit architecture by removing all MLP sub-blocks, adding
skip connections (to the MaxVit output) after each MaxVit sub-block,
and using normalized keys and queries in attention \citep{ViT-22B}.

Afterwards, we take the central crop of size 768 km by 768 km, and gradually upsample the
representation to 4 km resolution using skip connections from 
the downsampling path, at which point we again take a central crop, this time of size 512 km by 512 km.
The network outputs a categorical distribution over 256 bins for each of 6 ground weather variables
and a deterministic prediction for each of 617 assimilated weather state
channels using an MLP with one hidden layer applied to the representation at 4~km resolution.
For precipitation (both instantaneous rate and hourly accumulation),
we upsample the representation to 1~km resolution and
output for each pixel a categorical distribution over 512 bins.
Low-level details regarding the network architecture, optimization and hyperparameters used can be found in Supplement~B.

\begin{center}
   \centering
  \includegraphics[width=1\linewidth]{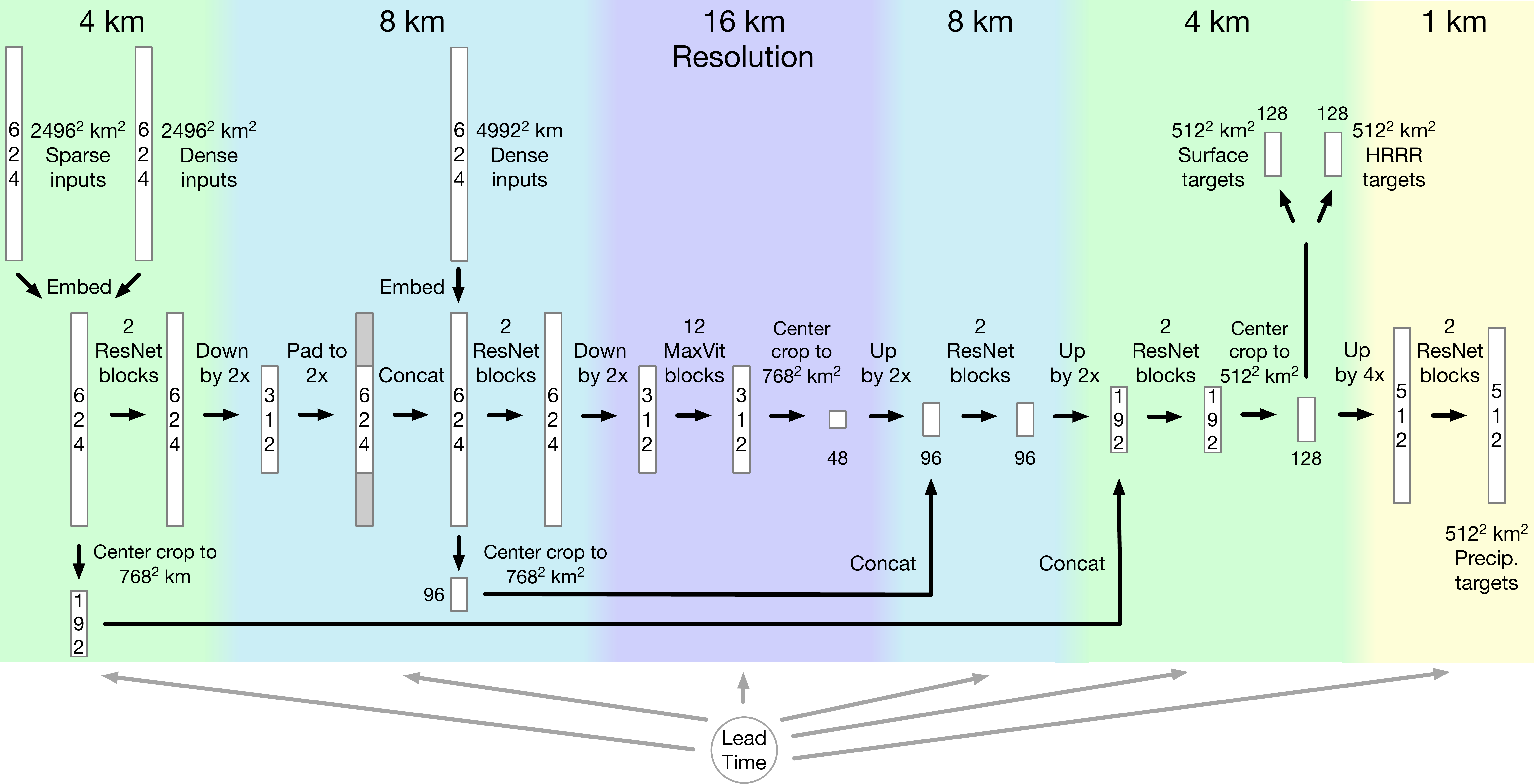}
\captionof{figure}{MetNet-3 network architecture.
Rectangles denote tensors and the numbers on/under them denote their spacial sizes in pixels.
}
\label{fig:models_flow}
\end{center}
\subsubsection{Conditioning with Lead Time}
Following MetNet-2~\cite{Espeholt2022}, we encode the lead time as a one-hot embedding with indices from 0 to 721 representing the range between 0 and 24 hours with a 2 min interval and map them into a continuous 32-dimensional representation. Instead of feeding the lead time embedding as an input, the embedding is applied both as an additive and multiplicative factor~\cite{perez2018film} to
the model inputs and to hidden representations before each
activation function or self attention block.
This ensures that the internal computation in the network
depends directly on lead time.

The task of forecasting weather becomes significantly harder as the lead time increases
which can negatively impact the model training.
To counteract it, we sample the lead time during training in a biased way
(exponential distribution) with $\texttt{t}=24\texttt{h}$ being sampled 10 times less frequently
than $\texttt{t}=0\texttt{h}$. We noticed that this sampling scheme improves the
results for all lead times including the long ones, which are sampled
less frequently.

\subsection{Training}

The network is trained to minimize the cross-entropy loss between the ground truth
data distribution and the model output.
For computational efficiency, the predictions for HRRR
assimilated state are deterministic and optimized with the Mean Squared Error (MSE) loss.
HRRR prediction loss is included in the model solely because it improves the quality
of the forecast for other variables
and we do not evaluate the predictions made by the model for the assimilated state.

\subsubsection{Densification}
While we only have the ground truth for surface variables at sparse locations,
the model needs to be able to generalize to all locations.
To this aim, we randomly mask out each OMO station with $25\%$ probability while training.
This ensures that the model is trained to predict OMO variables even
if there are no input OMO variables at the given location. (Note, this is separate from the $20\%$ hold-out set.)

We have also noticed that there is a trade-off between the quality of
precipitation and ground variables forecasts in a single model, and the results
can be slightly improved by having a separate model which specializes in predicting ground variables
but performs a bit worse for precipitation.
Therefore, we first train a model which is used for precipitation, and
afterwards we increase the weight of the OMO loss
by 100x compared to the precipitation model and finetune the model.
Moreover, we disable topographical embedding (fix them to zeros) for this OMO-specific model
because topographical embedding may hinder transfer between different locations, which
is crucial for learning only from targets present at a sparse set of locations.
See Figure~9 in Supplement~E for plots comparing the two models.

\subsubsection{Loss Scaling}

As the network is trained to optimize multiple losses
(cross entropy for instantaneous and accumulated precipitation rate as well as 6 OMO variables,
and MSE for 617 HRRR assimilation variables)
which may have very different magnitudes,
it is necessary to rescale them so that
their magnitudes are of similar order.
Apart from using standard techniques, namely rescaling all targets for the MSE loss so that each variable
has approximately mean $0$ and standard deviation $1$ and using manual scaling factors, we also
introduce a novel technique, which relies
on dynamically rescaling the gradient for each input-output sample.
More precisely, after calculating the gradient of the MSE loss w.r.t. the model output
for each sample, we rescale it, so that it has the same L1 norm for each output channel
without changing the overall magnitude of the gradient for the sample.
Let $g_{ijc}$ denote the spacial location $i, j$ and channel $c$ of the gradient
w.r.t model output, and $C$ denote the number of channels. We then use the following rescaled gradient instead of $g$:
\begin{align}
\hat{g}_{ijc} = \frac{C \cdot w_c}{\sum_{c'} w_{c'}} g_{ijc} &&
w_c = \frac{1}{\sum_{i'j'} |g_{i'j'c}|}
\end{align}
where the sums are over all channels ($c'$) and all spacial locations ($i', j'$) of the model output for a single
input-output sample.%
This scaling guarantees, that the influence of each output channel is bounded
and therefore even if a small fraction of the target channels are corrupted, their
effect on the model is limited.

\subsubsection{Hardware Configuration}

Due to large size of the input context and internal network representations
(2496~km by 2496~km at 4~km resolution and 4996 km by 4996 km at $8$ km resolution),
the network does not fit on a single TPU core.
Instead of reducing the resolution, which could negatively impact the forecast quality,
we use model parallelism.
We follow MetNet-2~\cite{Espeholt2022} and
split the inputs, internal representation and targets into a four by four grid processed by 16 interconnected TPU cores, with each TPU core responsible for 1/16 of the area.
The only exception to this rule is gridded attention in MaxVit, where we partition the data across TPU cores so that full attention windows are processed on a single core.
The necessary communication at each layer is handled automatically and efficiently~\cite{DBLP:journals/corr/abs-2105-04663,jax2018github}.

The network is trained on 512 TPUv3 cores, where each of the $32$ groups of $16$ TPU cores
process $2$ input-output samples and the gradients from each group are synchronously aggregated after processing each batch. The fully trained MetNet-3 model took 7 days to train.

\section*{Acknowledgements}

We would like to thank Marc van Zee for Flax code contributions, Stephan Rasp for reviewing and insightful discussions, Bill Myers and Daniel Rothenberg for discussions, Tyler Russell for data management and Thomas Turnbull for visualizations, as well as Jeremiah Harmsen, Carla Bromberg, Luke Barrington, Pramod Gupta, Aaron Bell and Jason Hickey for organizational contributions.

\newpage
\bibliography{main}
\bibliographystyle{plain}

\end{document}


\maketitle

\appendix

\section{Data}
\label{app:data}
\begin{figure}[h]
\centering
\begin{minipage}{.47\textwidth}
    \centering
    \includegraphics[width=\textwidth]{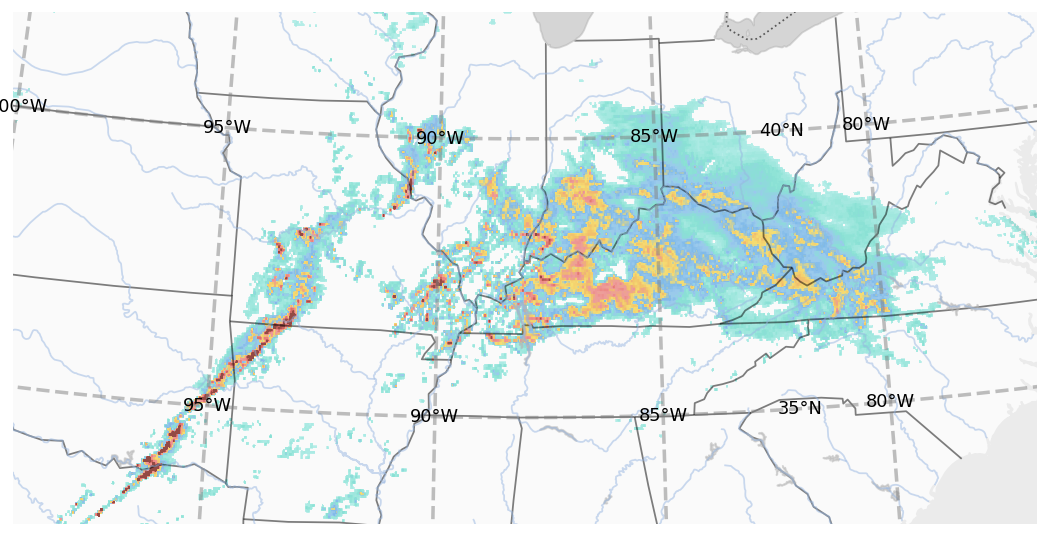}\\
     \end{minipage}%
    \begin{minipage}{.5\textwidth}   
    \centering
    \includegraphics[width=\textwidth]{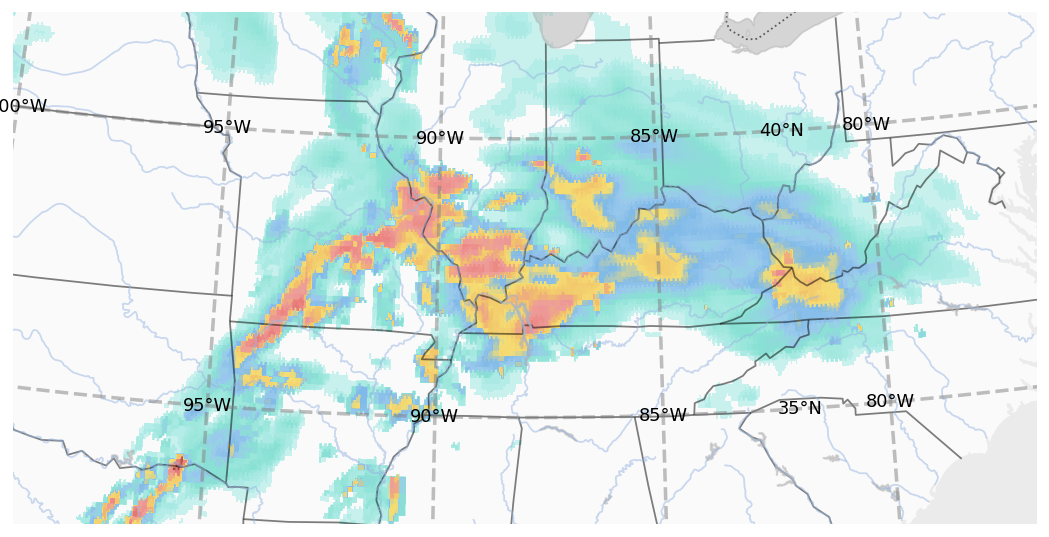}\\
    \end{minipage}\\
    \includegraphics[width=.3\textwidth]{assimilation_plots/dmi_colorbar.pdf}
    \caption{Precipitation rate in mm/h accordingly to MRMS (Left) and HRES assimilation (Right),
    for the timestamp Sat Nov 30 2019 12:00 UTC.}
    \label{fig:assimilation_hres_vs_mrms}
\end{figure}

\section{Supplement: Model and Training}\label{app:model_and_training}

Optimization hyperparameters can be found in Table~\ref{tab:optim}.
Below we list additional technical details related to the network architecture:

\paragraph{Inputs}

The high-resolution MRMS input has two channels --- instantaneous precipitation rate and precipitation type, while
the low-resolution MRMS input only contains the precipitation rate.
Precipitation rate inputs are preprocessed using the following transformation: $\tanh(\log(r+1)/4)$,
where $r$ is the precipitation rate in mm/h.
All other input channels are normalized to have mean and standard deviation values that are approximately 0 and 1, respectively.
We use time slices with the following offsets (in minutes) ---
high-resolution MRMS: -90, -75, -60, -45, -30, -25, -20, -15, -10, -5, 0;
OMO: -360, -180, -120, -60, -30, -15, -10, -5, 0;
all other inputs: 0.
Inputs are embeded to the internal representation of size 512 using a linear layer.

\paragraph{Network}
We use 512 channels throughout the whole network with the exception of 2 MLPs
(one at 4~km resolution and one at 1~km resolution)
which produce the network outputs which have a single hidden layer of size 4096.
All convolutions have kernels of size (3, 3) and are not dilated.
For computational efficiency, we use mixed precision \citep{mixed_precision}
with most of the computation performed in bfloat16 format.

\paragraph{MaxVit} We use 12 modified MaxVit \cite{maxvit} blocks.
   We introduced the following modifications compared to the original architecture:
   we removed MLP sub-blocks which were present in the original MaxVit architecture,
   we use normalized keys and queries \citep{Vit-22B} and we introduce skip connections
   from the output of each MaxVit block to the output of MaxVit.
   More precisely, the final output of MaxVit is a linear transformation
   of the outputs after each sub-block (after summing with the residual branch).
   All attention windows have size 8 by 8 and we use 32 attention heads.
   MBConv \cite{mbconv} in MaxVit uses the expansion rate of 4, and
   squeeze-and-excitation (SE, \cite{se}) with the bottleneck ratio of 0.25.

\paragraph{U-Net}
    In the downsampling path of U-Net, we apply 2 convolutional ResNet blocks and downsample by 2x with max pooling
   on 4 km and 8 km resolution levels.
   In the upsampling path, we upsample using a transposed convolution \citep{long2015fully}
   with kernel (2, 2) and stride (2, 2) on both 16 km and 8 km level, and then apply 2 convolutional
   ResNet blocks. Upsampling from 4 km to 1 km resolution is performed by repeating each activation
   across a 4 by 4 pixels square and applying again 2 ResNet blocks.
   \paragraph{Normalization} We use pre-activation (pre-LN, \cite{preLN}) layer normalization \citep{LN} throughout the network. We also apply layer normalization after each convolution which is not the last
   convolution in the given sub-block. 
   \paragraph{Lead Time Conditioning} We use additive-multiplicative conditioning (FiLM, \cite{film})
   on lead time throughout the network.
   The conditioning is applied to the the network inputs and after each layer normalization.
   All additive and multiplicative factors are outputted by a single
   MLP with one hidden layer of size 32 which takes as input one-hot encoded lead time.
   The second layer of this MLP is initialized so that at initialization the conditioning
   is an identity function.
   \paragraph{Topographical Embeddings} To limit the number of parameters in the topographical
   embeddings, we only allocate topographical embeddings
   for the region 14.8-59.9N, 150.7-39.3W.
   This results in 3M points on a grid with a
   stride of 4~km and 60M trainable parameters for embeddings of size 20.
   \paragraph{Activation Functions} We use GELU \cite{gelu} inside MBConv (in MaxVit) and ReLu in all other places.
   \paragraph{Initialization} We use LecunNormal initializer.
   Additionally we rescale the initialization of the last linear layer in each sub-block in MaxVit 
   by $1/\sqrt{N}$, where N is the number of sub-blocks those outputs are added on the given residual connection as
   described in \cite{child2019generating}.
   \paragraph{Regularization} We apply Dropout \cite{dropout} with the rate of 0.1 before adding the output
   of each sub-module to the residual branch and after the first convolution in each ResNet block.
   We use stochastic depth \cite{stochastic_depth} in MaxVit with the probability of dropping
   a given sub-module (i.e. MBConv, local attention or gridded attention) increasing linearly
   thorough the network from 0 to 0.2. We also use weight decay coefficient 0.1 as defined in AdamW \citep{adamW}.

 \begin{table}[h]
 \begin{tabular}{ l c } 
 \toprule
  \textbf{Training Hyperparameters} & \textbf{Value} \\
  \midrule
  Optimizer & AdamW \cite{adamW} \\
  Learning rate & 8e-5 \\
  AdamW $\beta_1$ & 0.9 \\
  AdamW $\beta_2$ & 0.999 \\
  Weight Decay & 0.1 \\
  Polyak Decay & 0.9999 \\
  Batch size & 64 \\
  Training steps & 260k \\
  OMO finetuning steps & 80k \\
  \bottomrule
 \caption{Optimization hyperparameters for MetNet-3.}
 \label{tab:optim}
 \end{tabular}
 \end{table}

\subsection{Outputs, Targets and Losses}

\begin{table}[H]
\begin{center}
\begin{tabular}{ l c  c c c c c} 
\toprule
 \textbf{Target} & \textbf{Resolution} & \textbf{\#Channels} & \textbf{Loss Function} & \textbf{\#Bins} & \textbf{Bin Size} \\ \midrule
    {MRMS rate} & 1 km & 1 & Cross Entropy & 512 & 0.2 mm/h \\
    {MRMS accumulation} & 1 km & 1 & Cross Entropy & 512 & 0.2 mm \\ \midrule
    {OMO temperature} & 4 km & 1 & Cross Entropy & 256 & 1 K \\
    {OMO dew point} & 4 km & 1 & Cross Entropy & 256 & 1 K \\
    {OMO wind speed} & 4 km & 1 & Cross Entropy & 256 & 0.1 knot \\
    {OMO wind components} & 4 km & 2 & Cross Entropy & 256 & 0.1 knot \\
    {OMO wind direction} & 4 km & 1 & Cross Entropy & 180 & 2 degrees \\ \midrule
    {HRRR assimilation} & 4 km & 617 & MSE & N/A & N/A \\
    \bottomrule
\end{tabular}
\caption{Details of outputs produced by MetNet-3.}
\label{tab:outputs_full}
\end{center}
\end{table}

Table~\ref{tab:outputs_full} lists different outputs produced by MetNet-3.
As the network is trained to optimize multiple losses,
which may have very different magnitudes,
it is necessary to rescale them so that
their magnitudes are of similar order of magnitude.
To this aim, we first rescale all targets for the MSE loss so that each variable
has approximately mean $0$ and standard deviation $1$,
and apply dynamic gradient rescaling described in the main article.

We also introduce additional manual scaling factors:
\begin{itemize}
    \item HRRR loss is multipled by 10 and divided by the number of HRRR channels being predict (617).
    \item We additionally increase the weight on HRRR channels corresponding to OMO ground variables the model predicts,
    namely 2m temperature, 2m dew point and 10m wind components by 30x. The weight of the remaining channels is decreased
    so that this step does not change the average weight of a HRRR channel.
    \item Each OMO target channel has the same weight with the sum of their weights being set to $0.01$ for the standard (precipitation) model
    and increased to $1$ for the OMO model finetuning.
\end{itemize}

\section{Supplement: Evaluation}
\label{sec:supp_evaluation}

\paragraph{Non-monotonic CRPS plots} We filter our evaluation dataset to only include locations and times when
historical forecasts for all baselines are available.
In particular, historical ENS forecasts are only available for two runs per day (00 and 12 UTC)
so all our evaluations only start at two times during the day.
Because of that, the expected amount of precipitation depends on the lead time.
Higher amounts of precipitation generally result in higher CRPS values, which results
in cases when CRPS counter-intuitively decreases with lead time.
We do not observe a similar phenomenon on CSI plots, because CSI scores
are by definition normalized (Supplement~\ref{sec:csi}).

\paragraph{MRMS and HRRR mask}
The quality of MRMS data varies between locations depending mostly on the distance from the nearest radar.
While we use all available data for training, we only evaluate using data from locations with the highest
quality of radar data (Figure~2).

\begin{figure}[H]
    \centering
    \begin{subfigure}{.48\textwidth}
      \centering
  \includegraphics[width=1\linewidth]{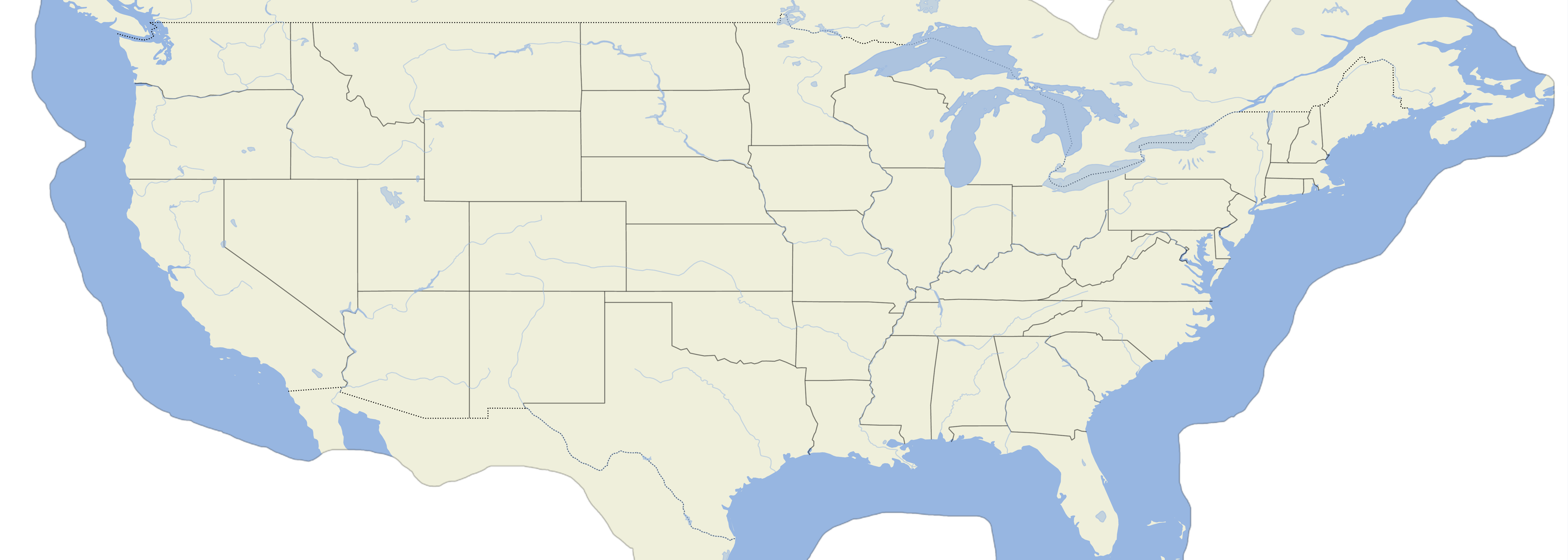}     
    \end{subfigure}
    \begin{subfigure}{.48\textwidth}
      \centering
  \includegraphics[width=1\linewidth]{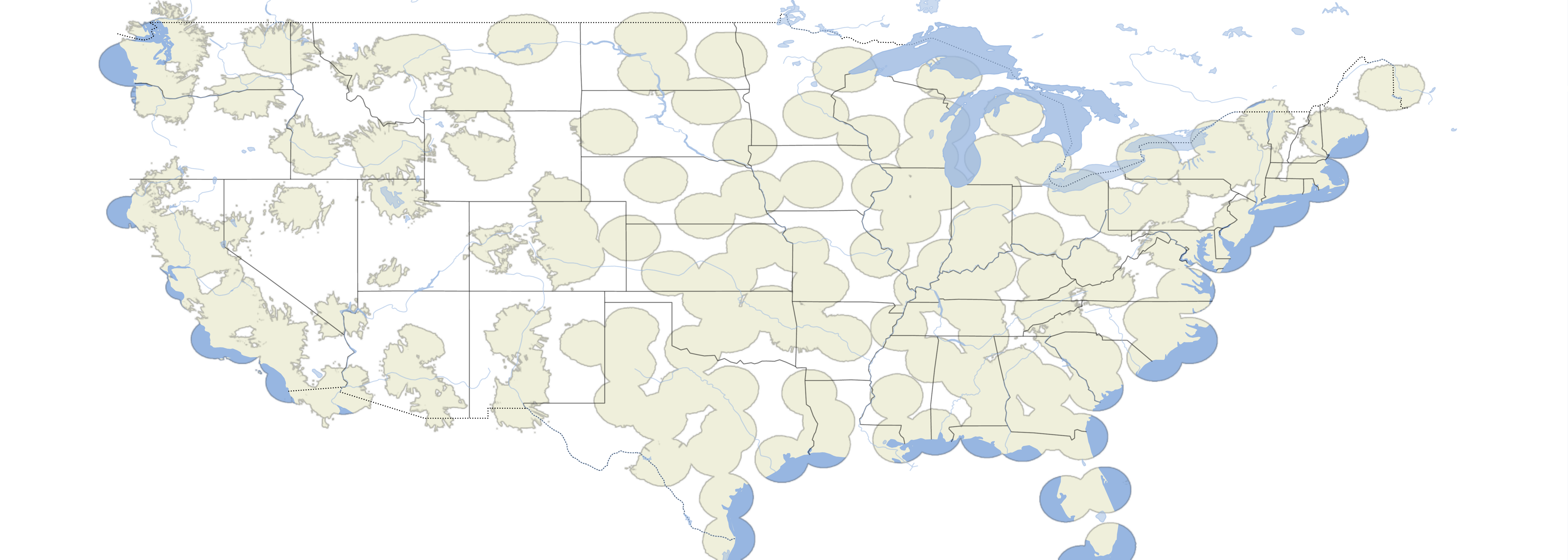}     
    \end{subfigure}
  \caption{Training (Left) and evaluation (Right) masks used for MRMS and HRRR targets.}
  \label{fig:mrms_masks}
\end{figure}

\paragraph{OMO ground truth} This figure represents the OMO network of weather stations, also known as 1-minute FAA Automated Surface Observing System (ASOS) or formerly high-frequency METAR.

\begin{center}
  \includegraphics[scale=.2]{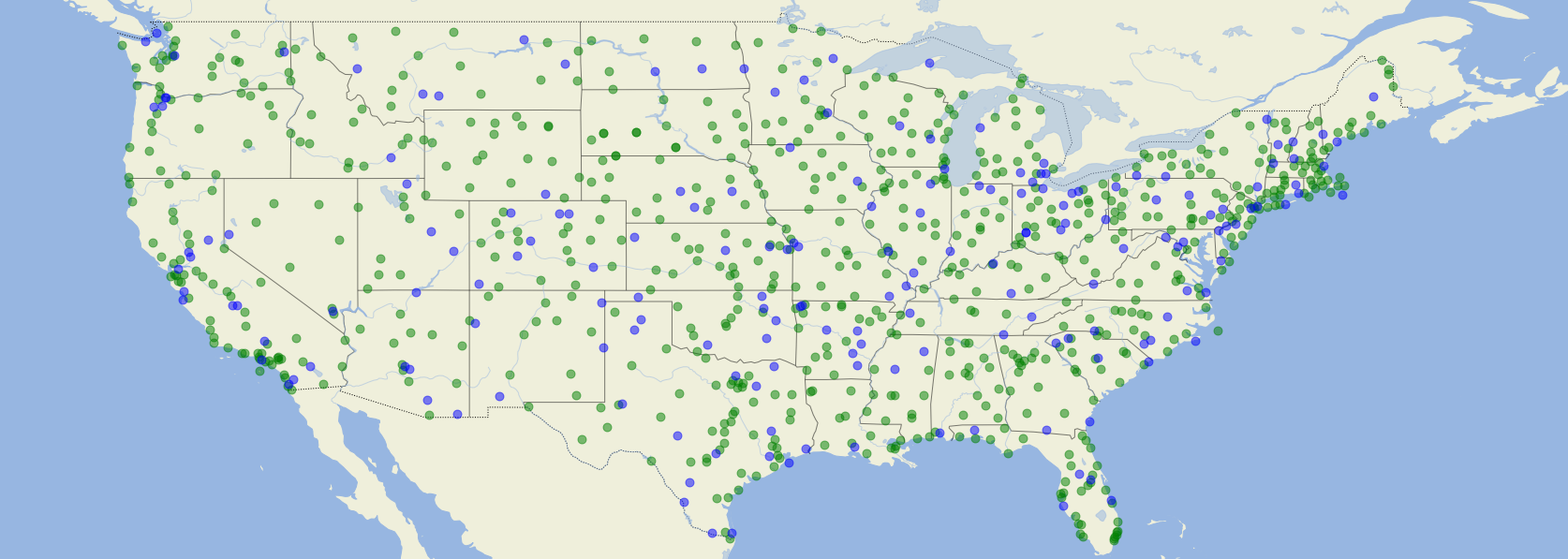}
  \captionof{figure}{OMO weather stations. Blue are test stations, and green are training stations.}\label{fig:metar_stations_map}%
\end{center}

\section{Supplement: Evaluation Metrics}\label{sec:metrics}
We evaluate the quality of the forecasts using three different metrics,
the Continuous Ranked Probability Score (CRPS)~\citep{DecompositionoftheContinuousRankedProbabilityScoreforEnsemblePredictionSystems},
the Critical Skill Index (CSI)~\citep{donaldson_rj_objective_1975}, 
and the Mean Absolute Error (MAE).

\subsection{Critical Success Index (CSI)}\label{sec:csi}

 The CSI score is a binary categorical score which we use
 to evaluate the quality of precipitation forecasts.
\begin{equation}
    CSI = TP/(TP+FN+FP)
\end{equation}
where TP are true positives, FN are false negatives and FP are false positives.
The CSI score is not directly applicable to the probability distributions that \mt{}
or ensemble baselines (HREF and ENS) produce.  To make a categorical decision, for a binary category corresponding to an amount of precipitation greater or equal to a given rate $r$, we calculate on a validation held-out set a probability threshold between~0~and~1
which maximizes CSI separately for each lead time.
If the total predicted probability mass for rates $\geq r$ exceeds the threshold, then we take it to be a positive prediction for this rate category. This is the same procedure as used in MetNet-2~\citep{Espeholt2022}.

We choose CSI over similar metrics for binary classification, because
 it disregards the number of true negatives, i.e. the cases when
 there was no precipitation (or the precipitation rate was below the specified evaluation rate)
 and the model predicted that correctly, and in the case of precipitation
 the vast majority of cases are of this type.

\subsection{Continuous Ranked Probability Score (CRPS)}
\label{sec:crps_metric}
CRPS in essence is the mean squared error between the cumulative density function (CDF) of the prediction and that of the ground truth integrated over the whole range of possible values.
We calculate it on the discretized set of values, i.e.
\begin{equation}
    CRPS = \sum_{i=1}^{N} (P_M(y \leq u_i) - \mathbbm{1}(y \leq u_i))^2 \times \texttt{bin size},
\end{equation}
where $i$ iterates over all discretization bins,
$u_i$ if the upper end of the $i$-th bin, $y$ is the ground truth
and $P_M(y \leq u_i)$ denotes the probability that $y \leq u_i$ under the model.

\subsection{Mean Absolute Error (MAE)}

Mean Absolute Error (MAE) is defined as
$$MAE = |\hat{y} - y|,$$
where $y$ if the ground truth and $\hat{y}$ is the deterministic prediction.
For MetNet-3 and ensemble baselines (HREF and ENS) we take
the median of the forecast distribution as $\hat{y}$.
We choose median, and not mean, because median minimizes MAE
for a perfect model.
MAE is not a suitable metric for very skewed distributions, and therefore we do no apply it
to precipitation.

\section{Supplement: Additional Results}\label{app:results}

\begin{figure}[h]
    \centering
    \begin{minipage}[t]{0.5\textwidth}
        \centering
        \textbf{(a) Instantaneous Precipitation Rate}\\
        \includegraphics[width=\textwidth]{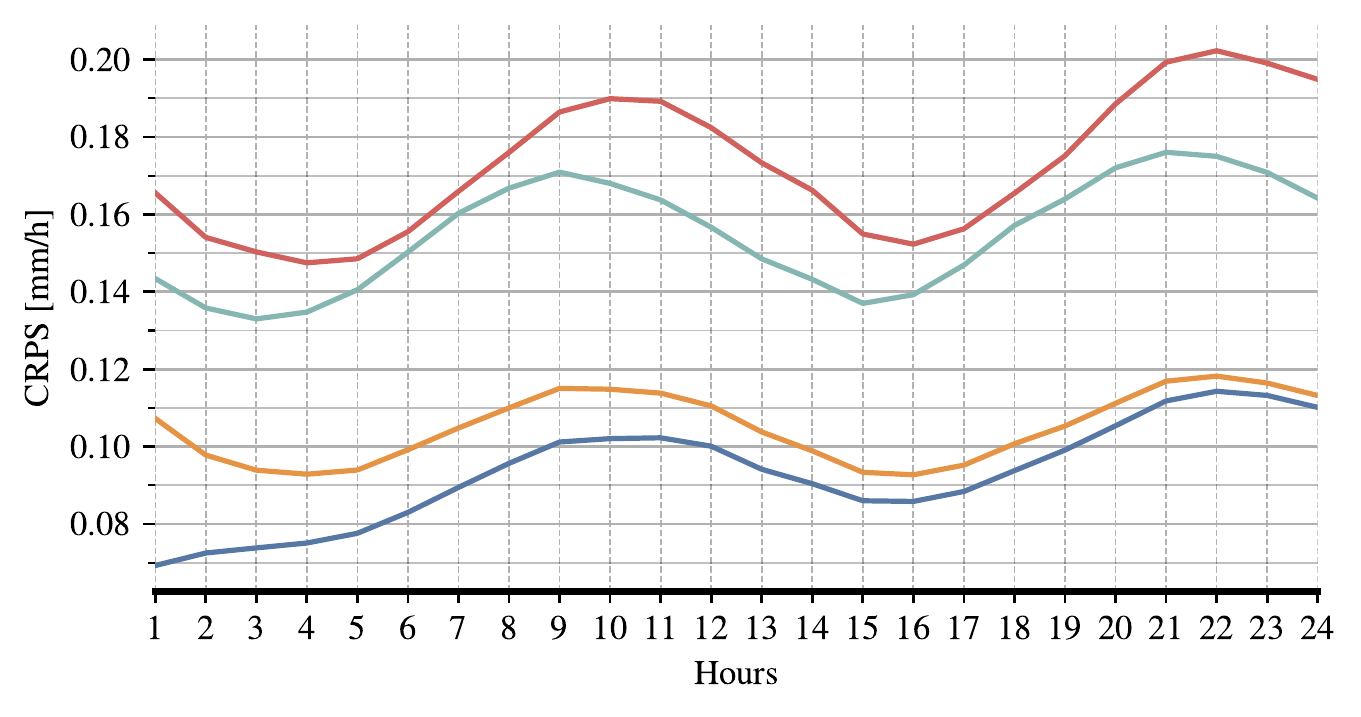}
    \end{minipage}%
    \begin{minipage}[t]{0.5\textwidth}
        \centering
        \textbf{(b) Hourly Accumulated Precipitation}\\
        \includegraphics[width=\textwidth]{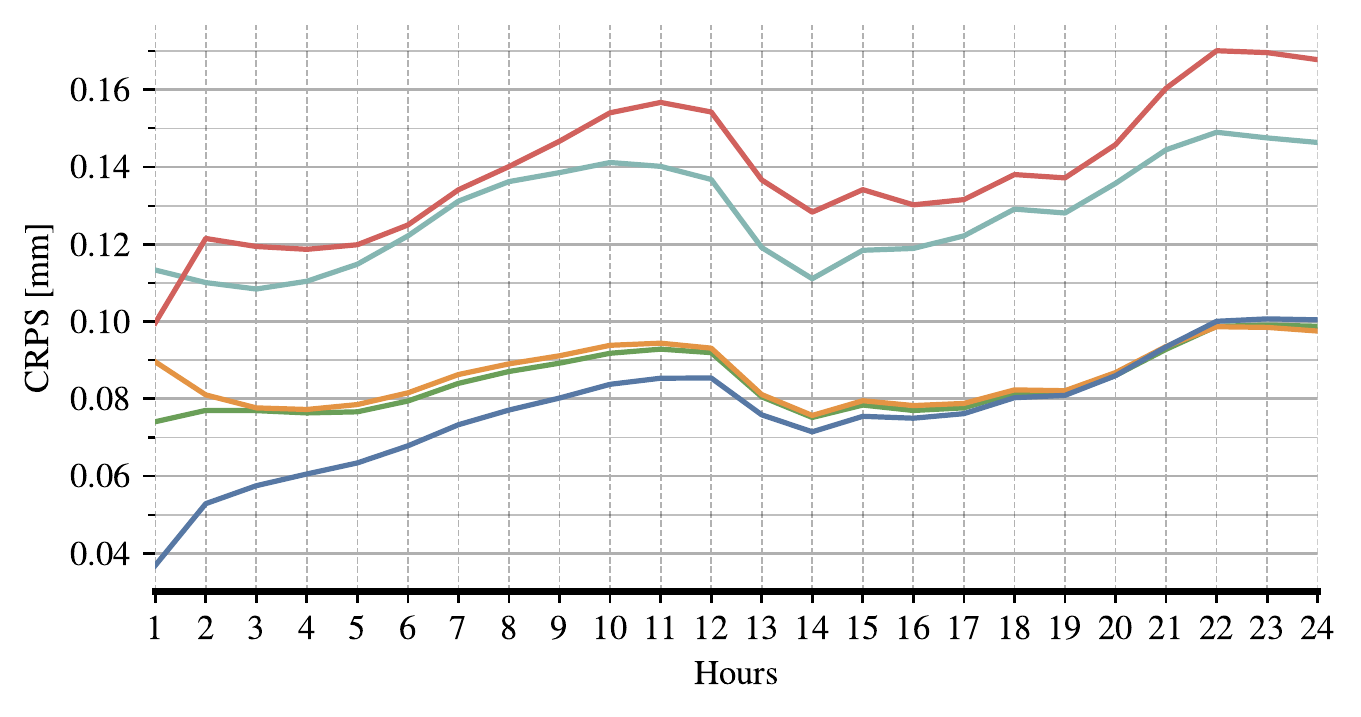}
    \end{minipage}
    \caption{CRPS values for MetNet-3 and baselines based on different precipitation measurements.}
    \label{fig:results_crps_all_baselines}
    \includegraphics[height=0.5cm]{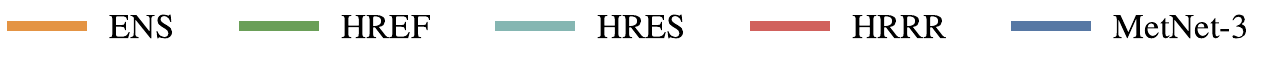}
\end{figure}

In this section we present some additional results:

\begin{itemize}
    \item Fig.~\ref{fig:results_crps_all_baselines}: CRPS plots for precipitation %
    including deterministic baselines (HRRR and HRES).
    \item Fig.~\ref{fig:csi_rate}: Instantaneous precipitation rate CSI plots for additional rates.
    \item Fig.~\ref{fig:csi_interpolated_cumulative}: Hourly accumulated precipitation CSI plots  for additional rates.
    \item Fig.~\ref{fig:results_windU}--\ref{fig:results_windV}:
    Results for surface wind U, V components.
    \item Fig.~9: Comparison of the standard version of MetNet-3 and the one finetuned
    for improved performance on ground variables.
    \item Fig.~\ref{fig:ablations}: Ablations with topographical embeddings and large-context inputs removed.
    \item Fig.~11: Comparison between MetNet-2 and MetNet-3.
\end{itemize}

\newcommand{\csifigure}[3]{
\begin{figure}[h]
\centering

\begin{minipage}{.5\textwidth}
\centering
\textbf{(a) 0.2 #3}\\
\includegraphics[width=\textwidth]{plots/#1__0.20mm__CSI.pdf}
\end{minipage}%
\begin{minipage}{.5\textwidth}
\centering
\textbf{(b) 1 #3}\\
\includegraphics[width=\textwidth]{plots/#1__1.00mm__CSI.pdf}
\end{minipage}\

\begin{minipage}{.5\textwidth}
\centering
\textbf{(c) 2 #3}\\
\includegraphics[width=\textwidth]{plots/#1__2.00mm__CSI.pdf}
\end{minipage}%
\begin{minipage}{.5\textwidth}
\centering
\textbf{(d) 4 #3}\\
\includegraphics[width=\textwidth]{plots/#1__4.00mm__CSI.pdf}
\end{minipage}\

\begin{minipage}{.5\textwidth}
\centering
\textbf{(e) 8 #3}\\
\includegraphics[width=\textwidth]{plots/#1__8.00mm__CSI.pdf}
\end{minipage}%
\begin{minipage}{.5\textwidth}
\centering
\textbf{(f) 20 #3}\\
\includegraphics[width=\textwidth]{plots/#1__20.00mm__CSI.pdf}
\end{minipage}\

\caption{CSI values for #2.}
\label{fig:csi_#1}
\includegraphics[height=0.5cm]{plots/#1__0.20mm__CSI_legend.pdf}\
\end{figure}
}

\csifigure{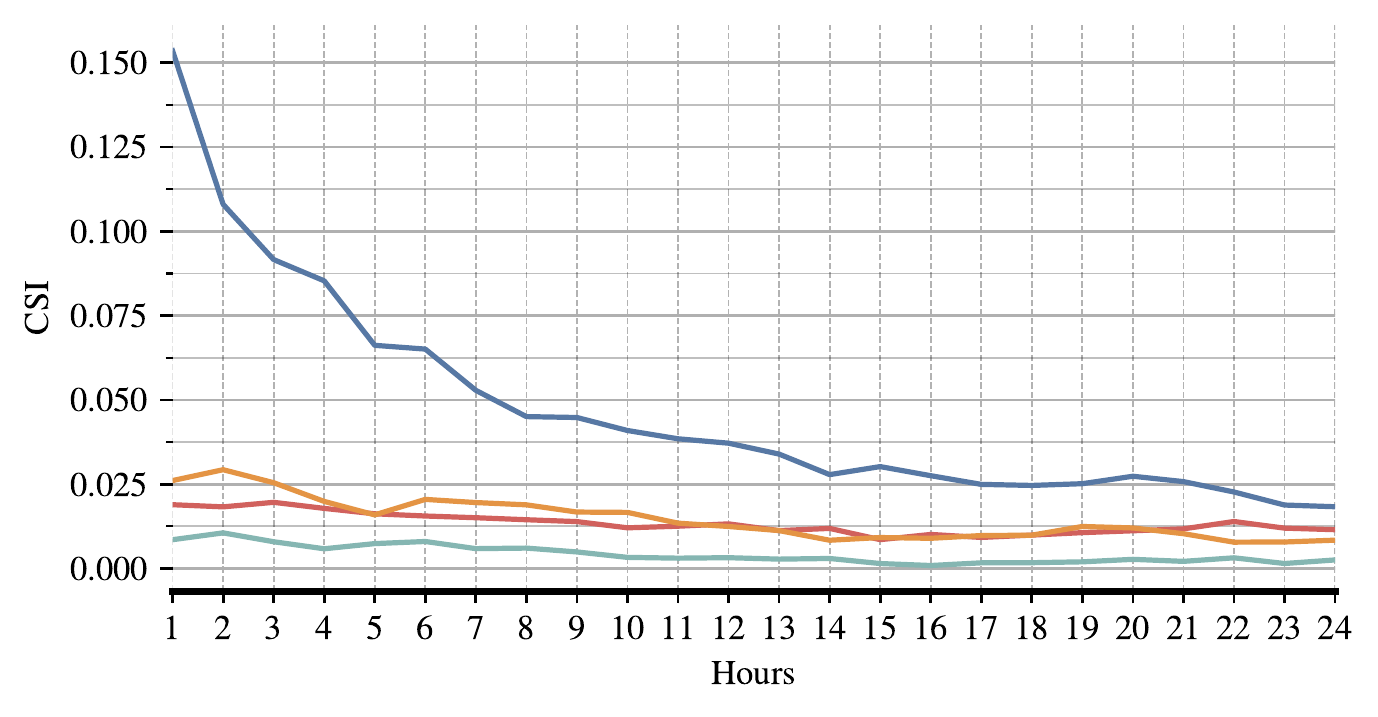}{instantaneous precipitation rate}{mm/h}
\csifigure{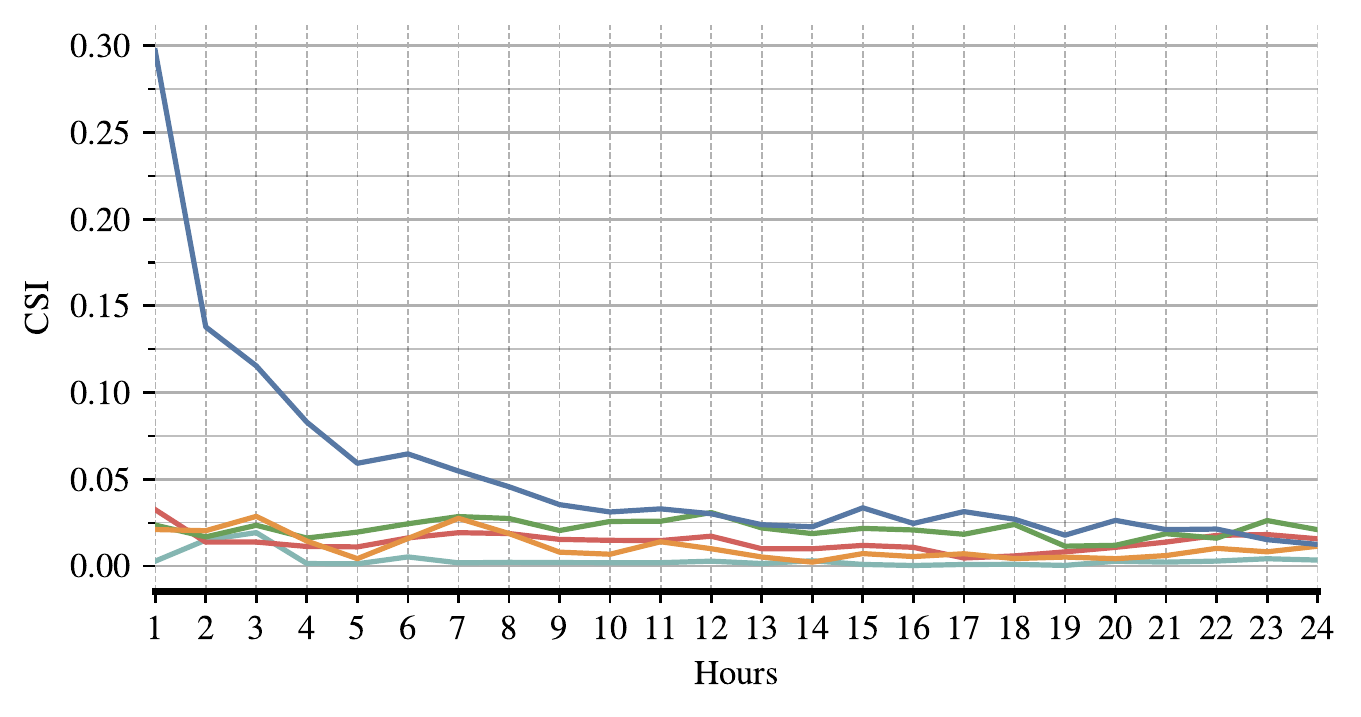}{hourly accumulated precipitation}{mm}

\newcommand{\metarplot}[2]{
\begin{figure}[h]
\centering
\begin{minipage}{.5\textwidth}
    \centering
    \textbf{(a) CRPS}\\
    \includegraphics[width=\textwidth]{plots/#1_CRPS.pdf}\\
\end{minipage}%
\begin{minipage}{.5\textwidth}   
    \centering
    \textbf{(b) MAE}\\
    \includegraphics[width=\textwidth]{plots/#1_MAE.pdf}\\
\end{minipage}\\
    \caption{Performance comparison between MetNet-3 and baselines for #2.}
    \label{fig:results_#1}
\includegraphics[height=0.5cm]{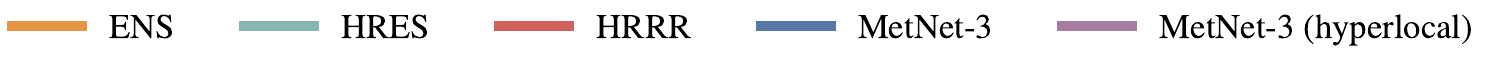}\\
\end{figure}
}
\metarplot{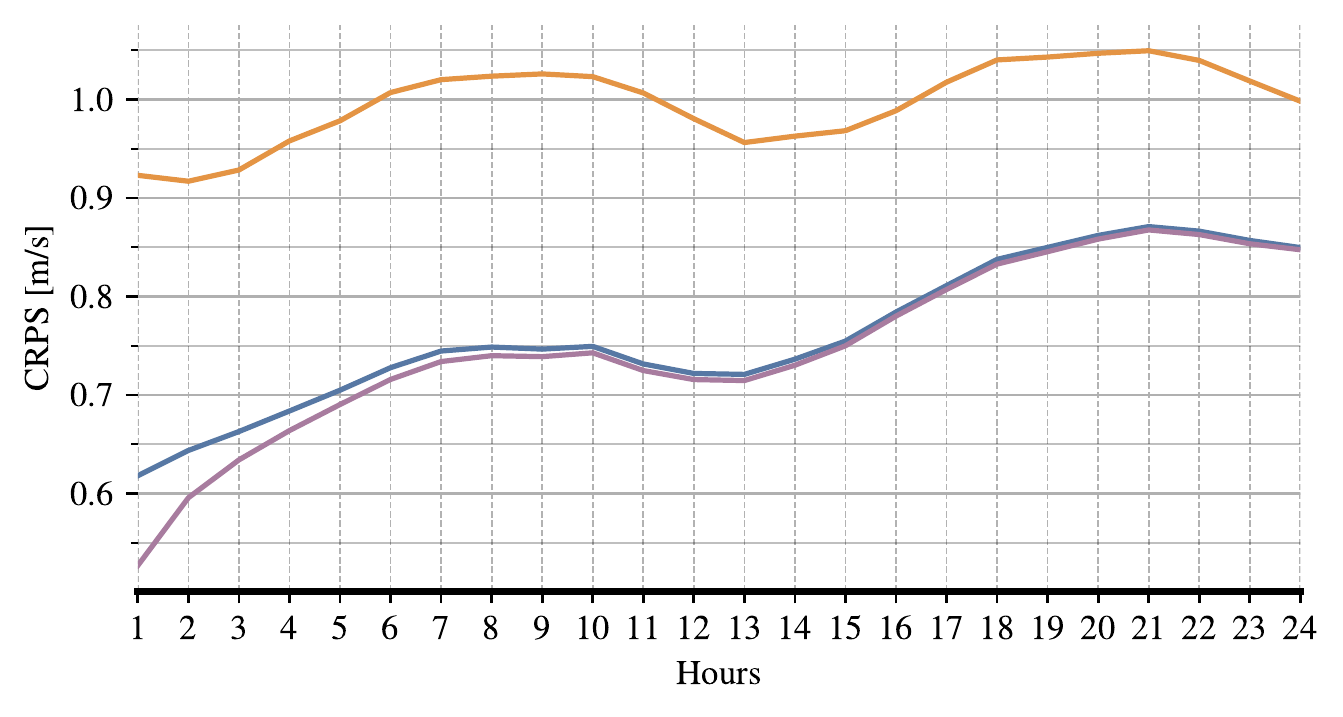}{U component of wind (i.e. eastward)}
\metarplot{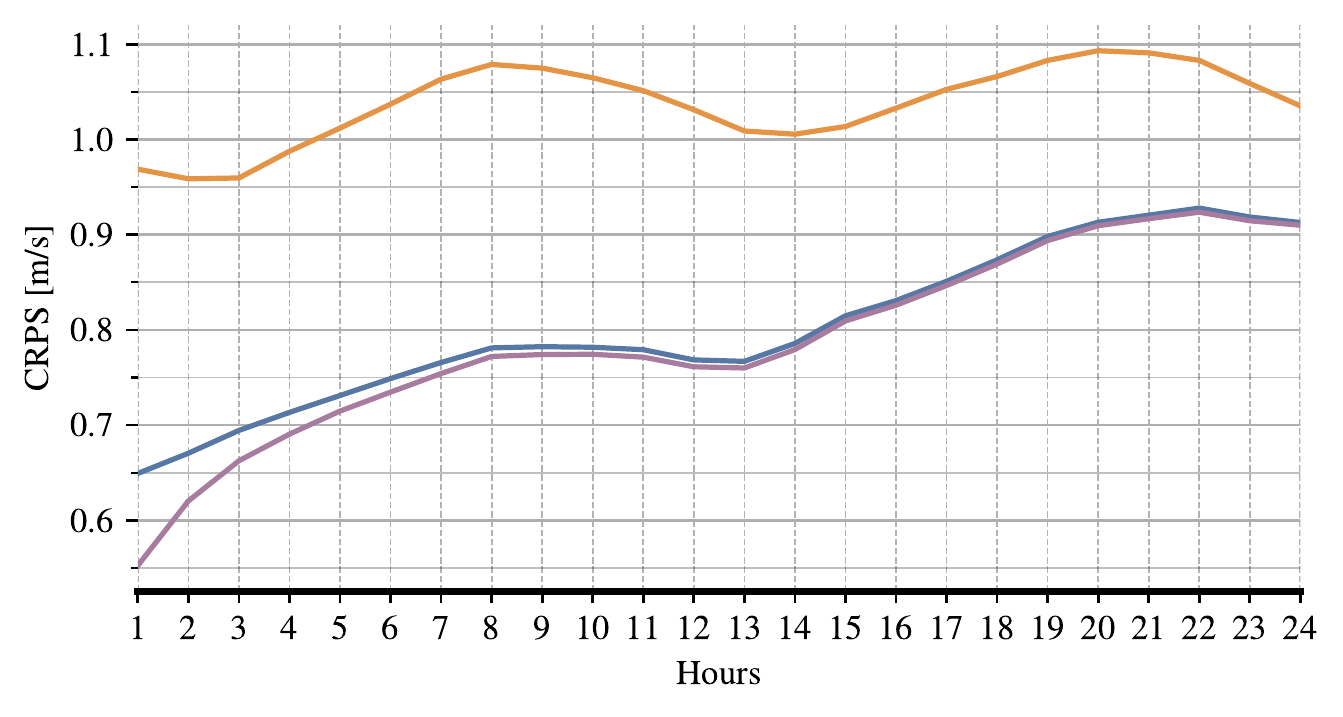}{V component of wind (i.e. northward)}

\begin{figure}[h]
\centering
\begin{minipage}{.5\textwidth}
    \centering
    \textbf{(a) Precipitation rate CSI 0.2 mm}\\
    \includegraphics[width=\textwidth]{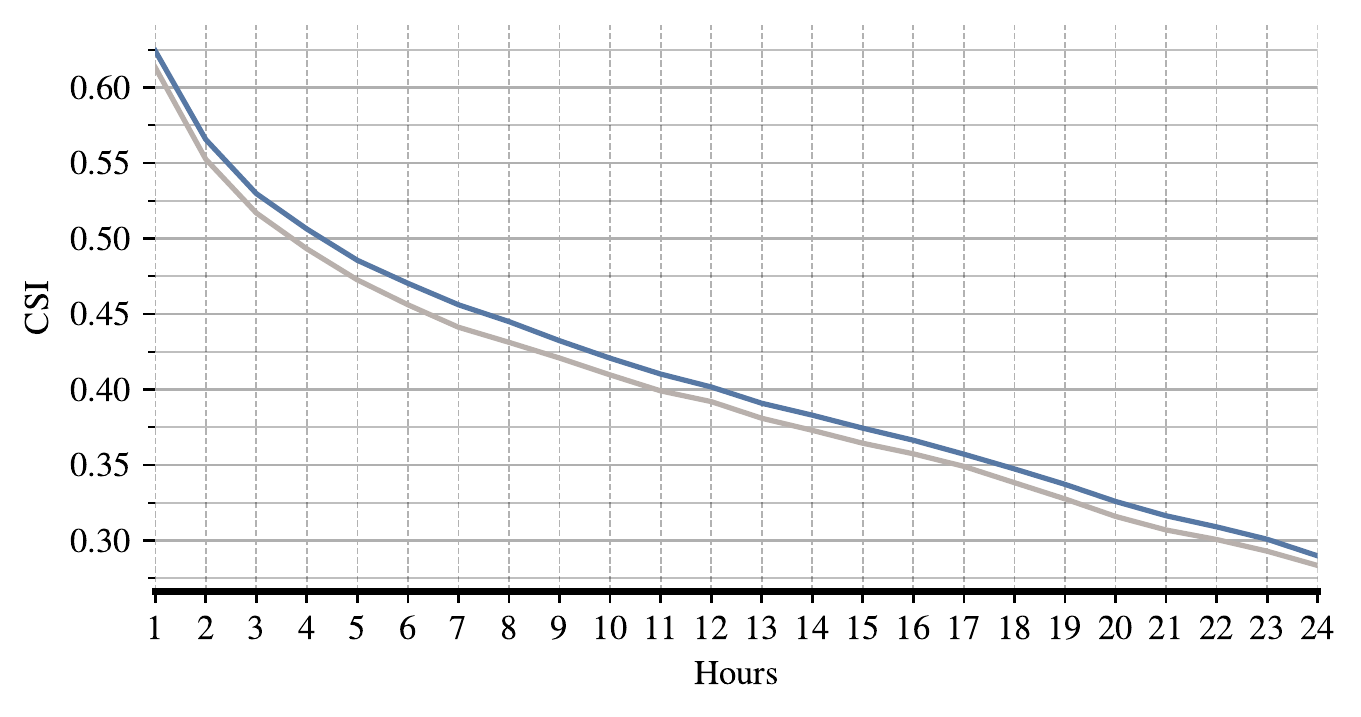}\\
\end{minipage}%
\begin{minipage}{.5\textwidth}   
    \centering
    \textbf{(b) Temperature MAE}\\
    \includegraphics[width=\textwidth]{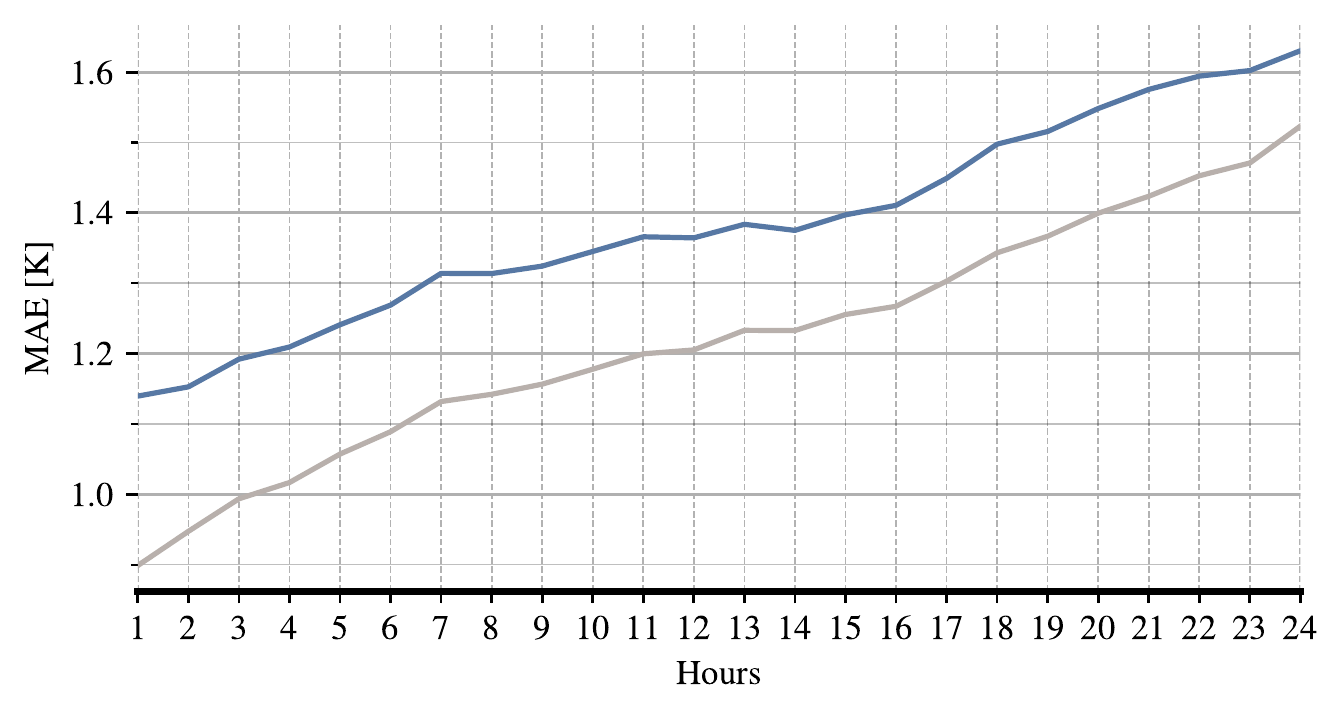}\\
\end{minipage}\\
    \caption{Effects of finetuning MetNet-3 for improved OMO performance.
    Notice that higher CSI and lower MAE are better.}
    \label{fig:finetuning}
\includegraphics[height=0.5cm]{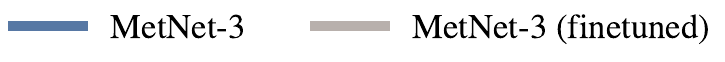}\\
\end{figure}

\begin{figure}[h]
    \centering
    \begin{minipage}[t]{0.5\textwidth}
        \centering
        \textbf{(a) CRPS}\\
        \includegraphics[width=\textwidth]{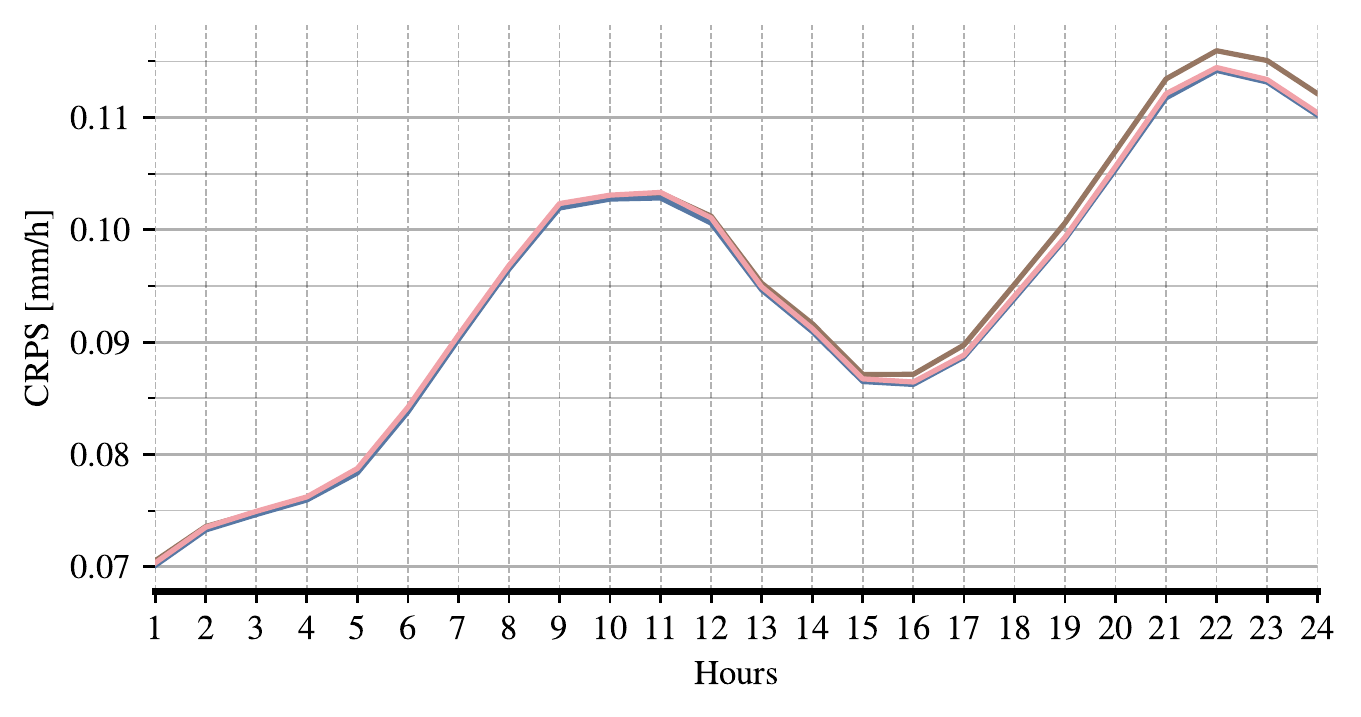}
    \end{minipage}%
    \begin{minipage}[t]{0.5\textwidth}
        \centering
        \textbf{(b) CSI 0.2 mm/h}\\
        \includegraphics[width=\textwidth]{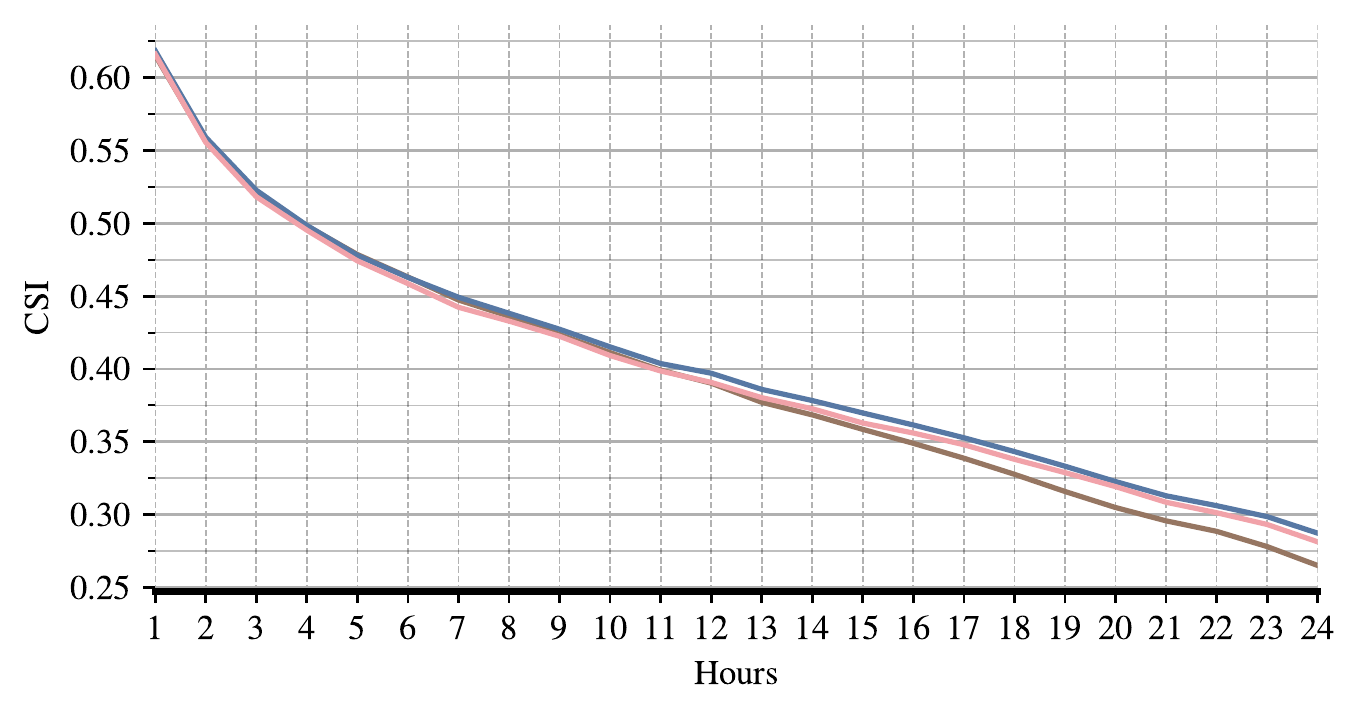}
    \end{minipage}
    \begin{minipage}[t]{0.5\textwidth}
        \centering
        \textbf{(c) CSI 1 mm/h}\\
        \includegraphics[width=\textwidth]{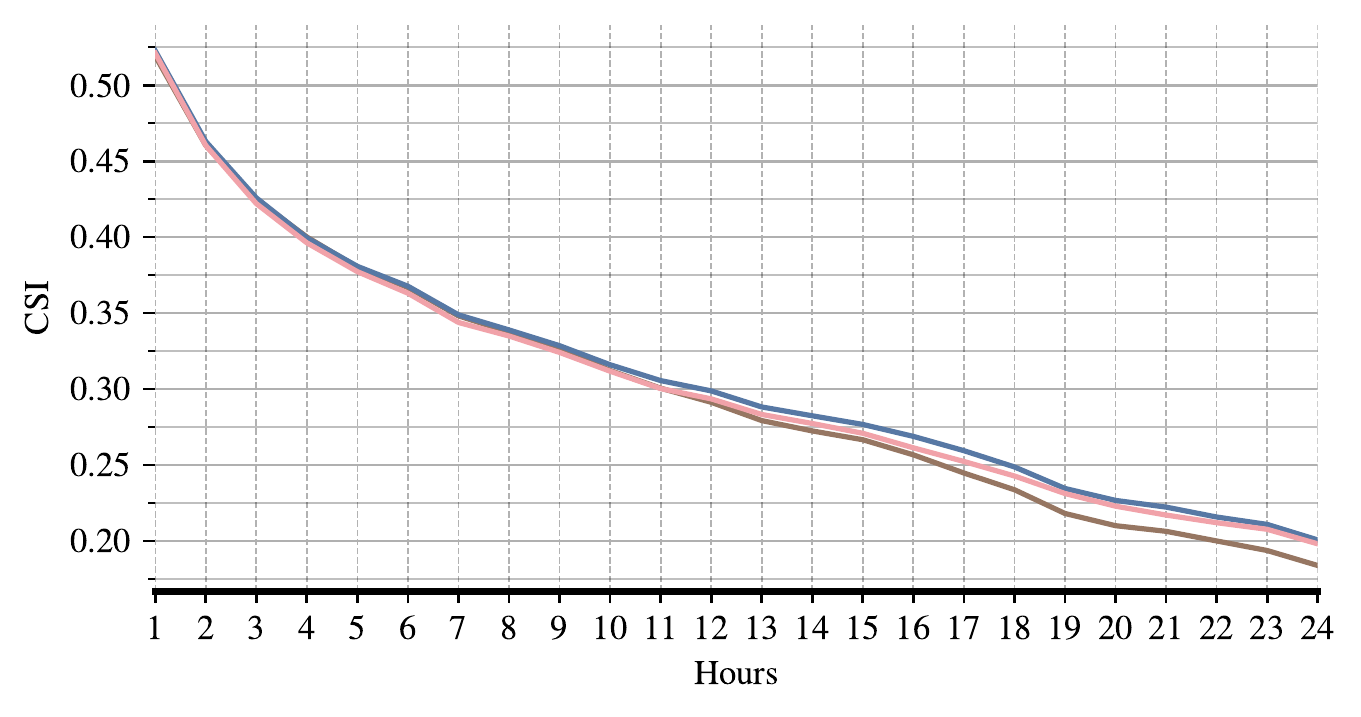}
    \end{minipage}%
    \begin{minipage}[t]{0.5\textwidth}
        \centering
        \textbf{(d) CSI 2 mm/h}\\
        \includegraphics[width=\textwidth]{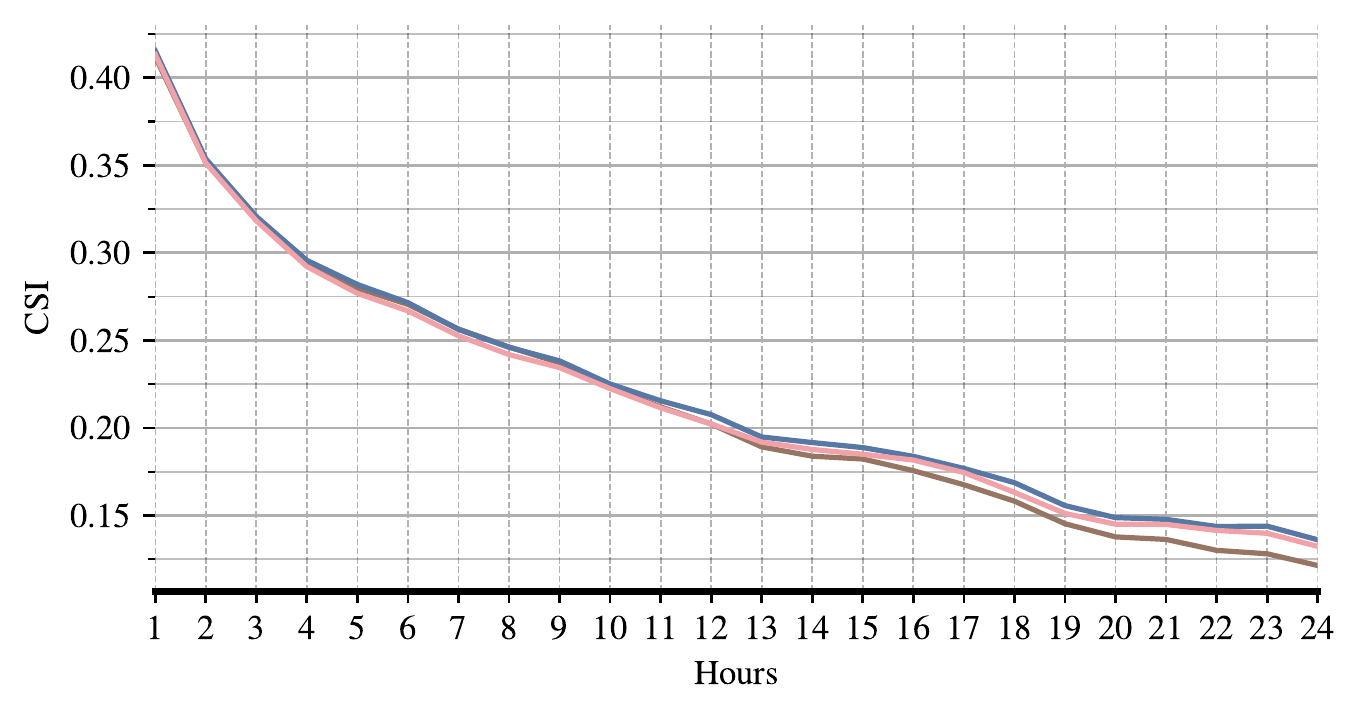}
    \end{minipage}
    \caption{Ablations with topographical embeddings and large-context inputs removed on
    instantaneous precipitation rate.
    All models shown in this figure have been trained for 150k steps.}
    \label{fig:ablations}
    \includegraphics[height=0.5cm]{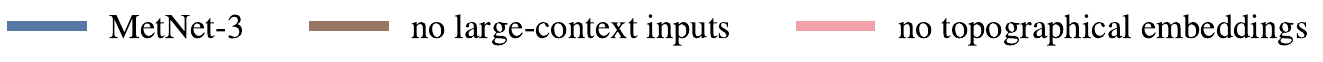}
\end{figure}

\begin{figure}[h]
    \centering
    \begin{minipage}[t]{0.5\textwidth}
        \centering
        \textbf{(a) CRPS}\\
        \includegraphics[width=\textwidth]{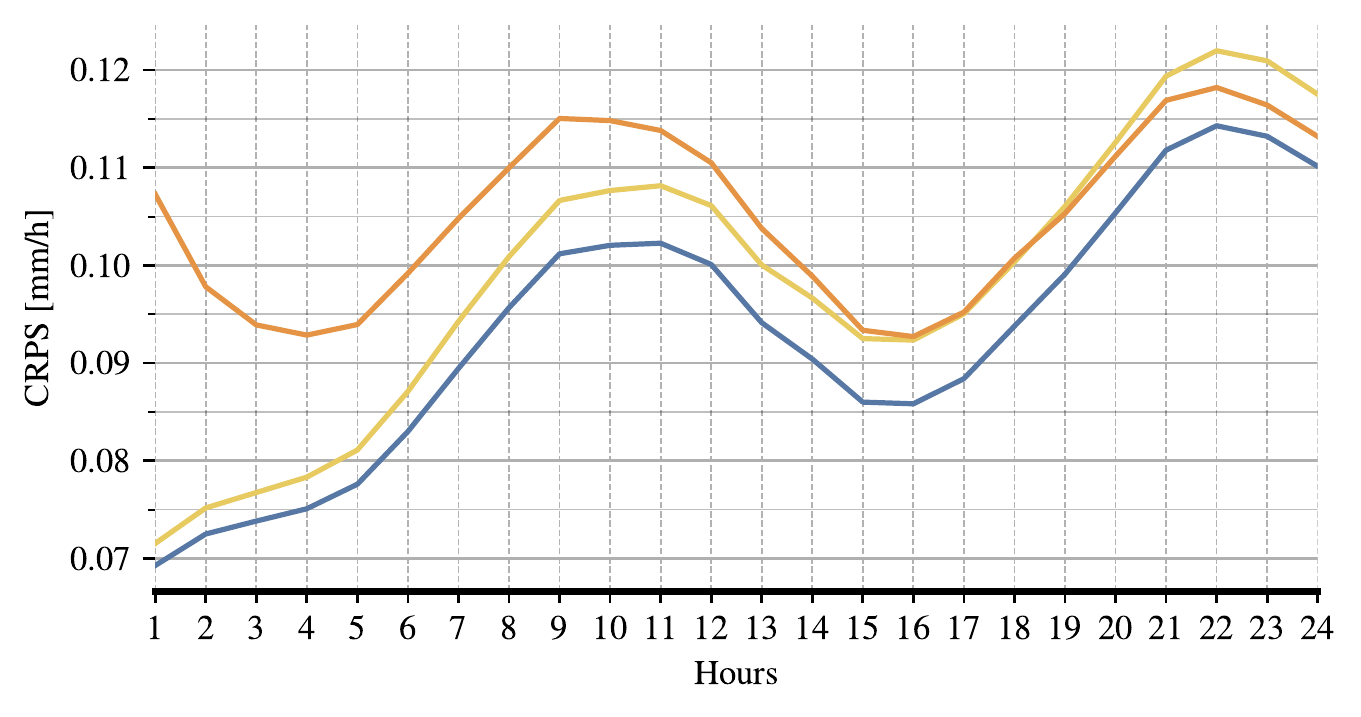}
        \label{fig:sub1}
    \end{minipage}%
    \begin{minipage}[t]{0.5\textwidth}
        \centering
        \textbf{(b) CSI 0.2 mm/h}\\
        \includegraphics[width=\textwidth]{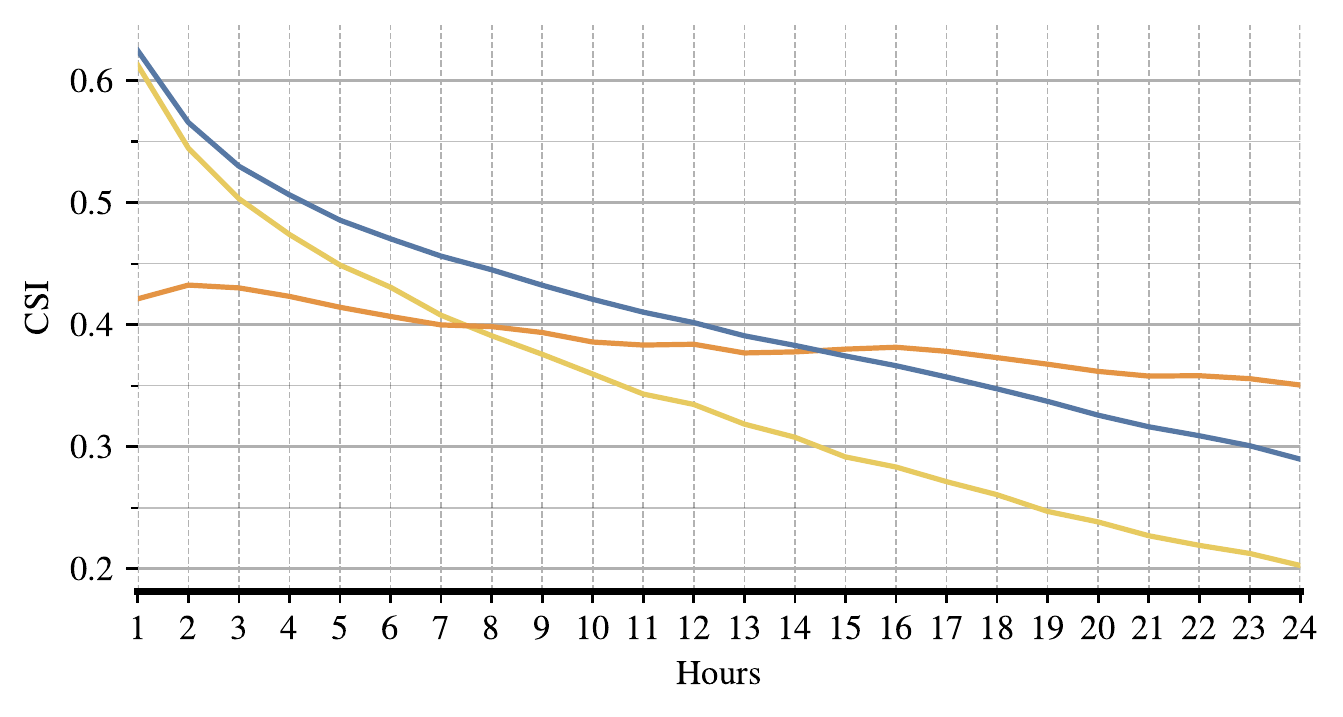}
        \label{fig:sub2}
    \end{minipage}
    \begin{minipage}[t]{0.5\textwidth}
        \centering
        \textbf{(c) CSI 1 mm/h}\\
        \includegraphics[width=\textwidth]{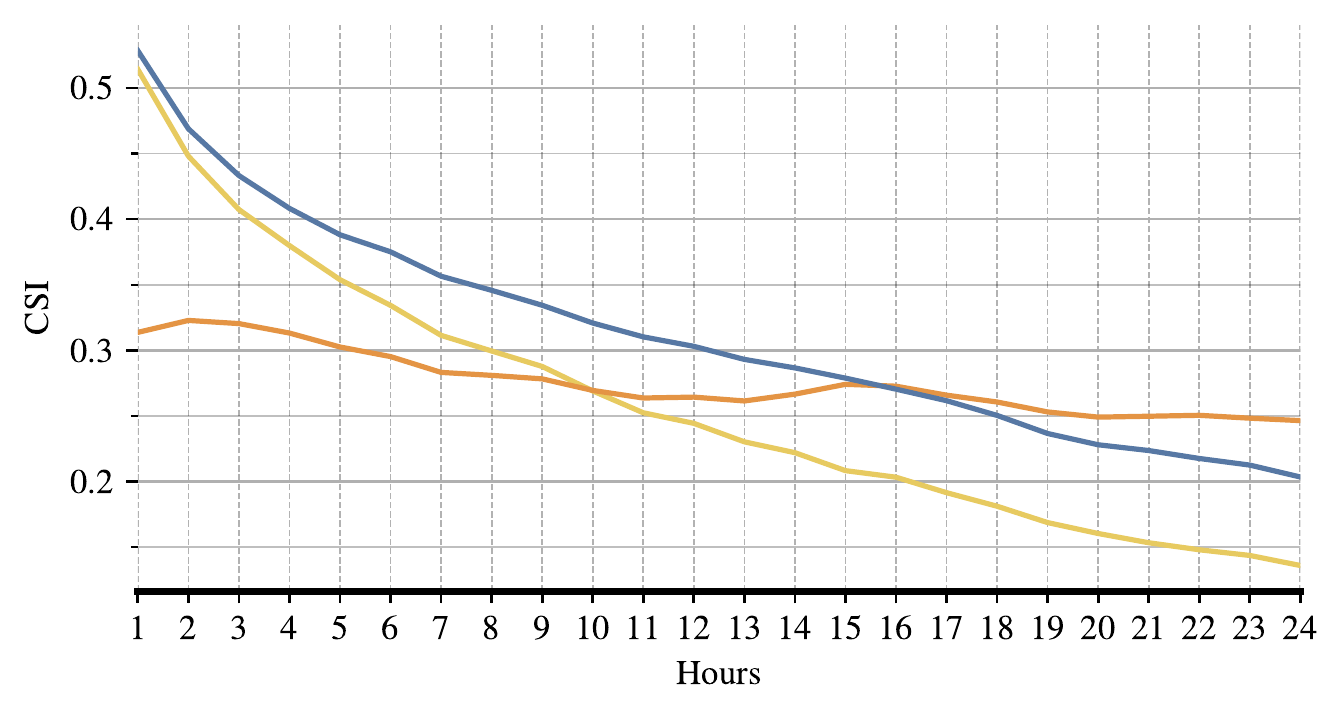}
        \label{fig:sub3}
    \end{minipage}%
    \begin{minipage}[t]{0.5\textwidth}
        \centering
        \textbf{(d) CSI 2 mm/h}\\
        \includegraphics[width=\textwidth]{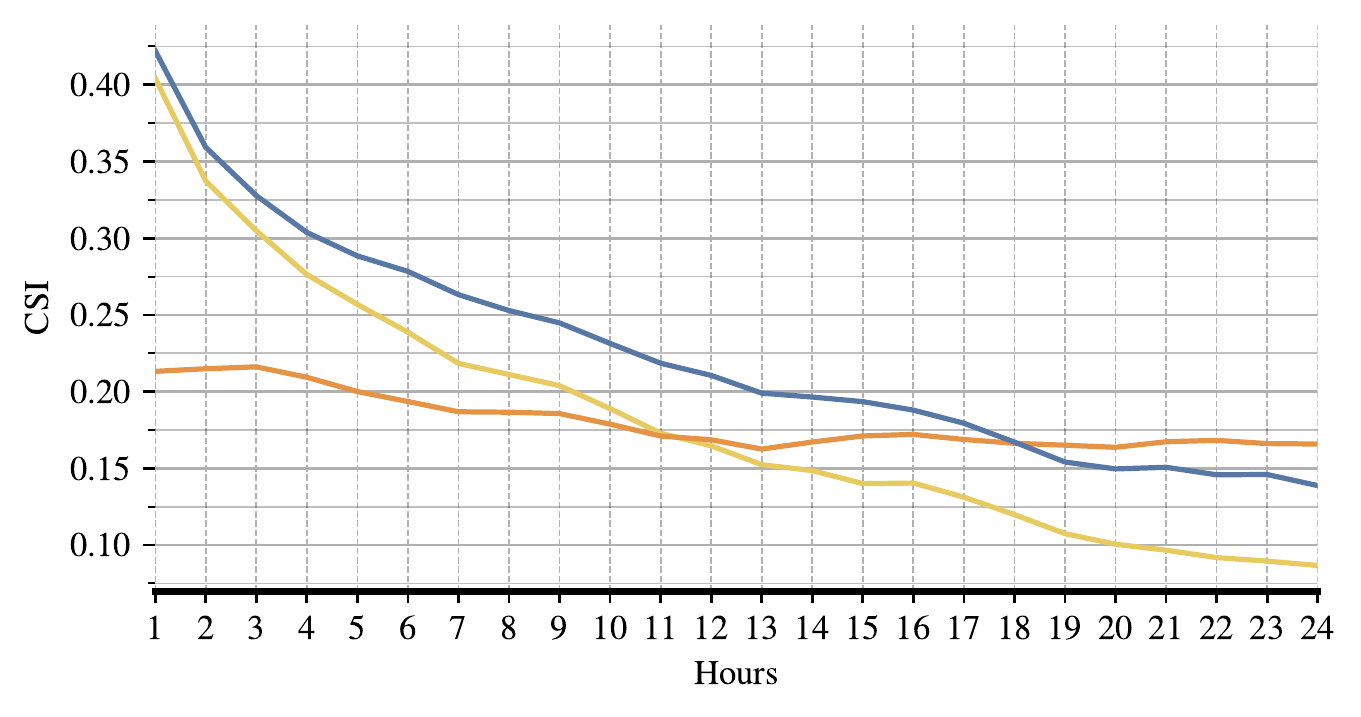}
        \label{fig:sub4}
    \end{minipage}
    \caption{Comparison with MetNet-2 on instantaneous precipitation rate.}
    \label{fig:results_metnet2}
    \includegraphics[height=0.5cm]{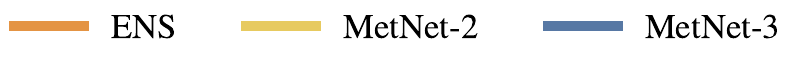}
\end{figure}

\bibliography{main}
\bibliographystyle{plain}